\documentstyle[12pt,psfig,aps]{revtex}
\def\baselinestretch{1.5}
\begin{document}
\pagenumbering{arabic}
\title{\bf  Relativistic Hartree approach including \\   
            both positive- and negative-energy  bound states} 
\author{G.~Mao, H.~St\"{o}cker, W.~Greiner}  
\address{Institut f\"{u}r Theoretische Physik der
J. W. Goethe-Universit\"{a}t \\
Postfach 11 19 32,  D-60054 Frankfurt am Main, Germany}
\date{\today}
\maketitle
\begin{abstract}
\begin{sloppypar}
We develop a relativistic model to describe the bound states of positive
energy and negative
energy in finite nuclei at the same time. Instead of searching for the negative-energy 
solution of the nucleon's 
Dirac equation, we solve the Dirac equations for the nucleon and 
the anti-nucleon simultaneously. The 
single-particle energies of negative-energy nucleons are obtained through
changing the sign of the single-particle energies of positive-energy 
anti-nucleons.
The contributions of the Dirac sea to the source terms of the
meson fields are evaluated by means of the derivative expansion up to the   
leading derivative order for the one-meson loop and one-nucleon loop.
After refitting the parameters of the model to the properties of spherical
nuclei, the results of positive-energy sector are similar to that calculated 
within the commonly used relativistic mean field theory under the no-sea
approximation. However, the bound levels of negative-energy nucleons vary
drastically when the vacuum contributions are taken into account. It implies
that the negative-energy spectra deserve a sensitive probe to the effective
interactions in addition to the positive-energy spectra.

\end{sloppypar}
\bigskip
\noindent {\bf PACS} number(s): 21.60.-n; 21.10.-k                       
\end{abstract}
\newcounter{cms}
\setlength{\unitlength}{1mm}
\newpage
\begin{center}
{\bf I. INTRODUCTION}
\end{center}
\begin{sloppypar}
The extension of the periodic system has a long history in chemistry and
physics. Particularly at the end of the sixties and the beginning of the 
seventies, the time when heavy-ion collisions had its advent, the idea of
super-heavy elements emerged. Besides the extension of the periodic system
to the {\em islands of super-heavy nuclei} within the proton $Z$- and neutron
$N$-axis, there exist two fascinating direction, i.e., to extend 
$Z$ and $N$ into the negative sector, $\bar{Z}$ and $\bar{N}$, and into
the multistrangeness dimension.  The first idea
was put forward in detail in Ref. \cite{Gre95}
where a {\em collective anti-matter production mechanism} was proposed. The main
physical considerations are the following: In the relativistic treatment of 
nuclear phenomena, one uses the Dirac equation to describe nucleons in nuclei
 \begin{equation}
\left[ \mbox{\boldmath $\alpha$} \cdot {\bf p} + \beta (M_{N} + S) + V \right]
 \Psi =E\Psi. 
 \end{equation}
Here $S$ and $V$ are the scalar and time-like vector potentials acting on 
nucleons in the nuclear medium. 
In the simplest version of relativistic mean field theory the scalar and vector
interaction are mediated by the sigma- and omega-meson exchange, respectively.
The Dirac equation has two solutions, i.e., the
positive-energy solution and the negative-energy solution. For the positive-energy
solution, the nucleon central potential $U_{cen} \sim V+S$ while for the 
negative-energy solution, due to the ``G-parity''
\footnote{The ``G-parity'' is an internal symmetry of the exchanged mesons in
strong interactions, which connects the $NN$ potential $U(NN)=
\sum_{m} V_{m}$
to the $N\bar{N}$ potential $U(N\bar{N})=\sum_{m} G(m)V_{m}$. $G$ is the 
combination of the isospin symmetry and the $C$-conjugation rule and is 
usually defined as $G=C \exp (-i\pi I_{2})$. In particular, $G(\sigma)
=G(\rho)=1$ and $G(\omega)=G(\pi)=-1$.},
the central anti-matter potential becomes
$\bar{U}_{cen} \sim -V+S$. Relativistic mean-field (RMF) calculations
in nuclear matter predict $V\sim 300$ MeV and $S \sim -350$ MeV \cite{Ser86}. 
In these calculations the vacuum contributions have been neglected and the 
definite values of $V$ and $S$ are parameter-dependent. If these two 
values for $V$ and $S$ are inserted into the formulae of the central potentials,
one immediately obtains
 $U_{cen} \sim -50$ MeV and $\bar{U}_{cen} \sim -650$ MeV. Therefore,      
the bound states of negative energy might be much deeper than the bound 
states of positive energy, although the exact potential 
depth of the bound states of
negative energy  has not been experimentally verified
 up to now. Furthermore, the
vector potential $V$ increases linearly with increasing of density. In violent 
relativistic heavy-ion collisions, the density may become very large,
$\rho \sim 5-10 \rho_{0}$. The potential of
 nucleons of negative energy (i.e., the anti-nucleons), may become larger than
twice the nucleon's free mass with increasing density. Then, 
 nucleons may be spontaneously emitted \cite{Mis93}. But nucleons will
also be emitted due to the dynamics, i.e., due to the time dependent change of
their orbitals, i.e., due to 
the Fourier frequencies of the
time dependent potentials and complex scatterings in the
 violent compression processes of a heavy-ion collision, and also due to
temperature built-up in such violent encounters.
These can create a great number of nucleon holes (i.e., anti-nucleons) in bound
states. Then, ''clusters'' of holes (anti-nucleons) are distributed over
shell-model-like orbitals. They are bound  by the meson fields created by the 
compressed matter. They are, so to speak,
 prepared collectively  for fusion into 
anti-matter clusters. Again, due to the Fourier  frequencies of the violent
nucleus-nucleus dynamics during and after the encounter, 
these anti-matter
clusters can be kicked into the anti-cluster continuum, i.e., the 
negative-energy continuum, and may escape collectively out of the reaction
zone. Another possible mechanism for anti-matter cluster escaping from
the reaction zone is due to the space-time dependent event by event
fluctuations of $N$-body phase space 
of anti-nucleons which
is not a smooth distribution function as in the case of $1$-body
calculations. 
Such collective creation processes of anti-matter clusters have a  high
potential to increase the probability for the production of light  
anti-matter clusters than the standard processes of particle scattering
and coalescence.

To realize the above theoretical conjecture for the production of anti-nuclei,
-- and analogously also for multi-$\Lambda$, multi-$\bar{\Lambda}$ nuclei, 
one should answer at least three questions in a rigorous way both from the
theoretical side and the experimental side: Firstly, how deep are the bound states
of negative energy (i.e., the bound states of anti-nucleons) in nuclei? 
Secondly, how does the potential
of anti-nucleons increase with the increase of density during the violent
relativistic heavy-ion collisions and what is the dynamical production procedure
 for the anti-matter clusters? Thirdly, can we really have stable 
or meta-stable (heavy)
anti-nuclei? In this work, we build a model to deal with the first question,
which might be extended to answer the third question as well in the future.

It is well known that relativistic mean-field theory had great success
in describing the ground states of finite nuclei, i.e., the bound states of
positive energy \cite{Rei86,Gam90,Bog81,Ren96}. In this model, one usually 
 considers only the
positive-energy solution of the Dirac equation and neglects the contributions
from the vacuum (the so-called {\em no sea approximation}). With the parameters
obtained as providing the {\em best-fit} to the properties of spherical nuclei,
such as NL1 \cite{Rei86} and TM1 \cite{Sug94} parameter sets, the model 
describes well the ground-state properties of spherical nuclei as well as the
deformation properties of even-even nuclei. 
Extensive investigation of various parameter sets in connection with 
predicting shells for superheavy nuclei have recently been given in Ref. 
\cite{Rut97}. However, this all has been done in the no-sea approximation
and without vacuum corrections.
The later, the effects of quantum
corrections, i.e., the vacuum contributions and their effects 
on the bound states of positive
energy were investigated by several authors \cite{Hor84,Per86,Was88,Fox89,Fur89}
within the local density approximation for the one-nucleon loop, the one-meson
loop and for the  derivative expansion for the one-nucleon loop only 
(in a chiral model,
loop corrections have been investigated in nuclear matter e.g., in Ref. 
\cite{Ain88}). 
The relativistic
mean-field approximation is extended to the relativistic Hartree approximation
(RHA). In that version, i.e., RHA, the parameters are fitted to the saturation
properties of nuclear matter as well as the rms charge radius in $^{40}$Ca.
The {\em best-fit} procedure within RHA to the properties of spherical nuclei
has not been performed yet. In these preliminary studies it is found that 
 the vacuum contributions do not improve the 
systematics of nuclei over RMF but the scalar density is decreased in the 
 interior
of nuclei. It, in turn,  will cause the decrease of the scalar field $S$ 
because the scalar field is coupled to the scalar density. Correspondingly,
the vector field $V$ will also decrease since the value 
$V+S$ is controlled by the
saturation properties. Therefore, even if the vacuum corrections may not cause 
significant difference to the bound states of positive energy after refitting
the parameters, they should have strong influence on the bound states 
of negative energy which are sensitive to the sum of scalar and vector  
field $-V+S$ rather than the cancellation $V+S$ (remember: $V$ is positive,
$S$ is negative!).

In the present paper we will develop a relativistic Hartree approach for the
bound states of positive energy and negative energy in finite nuclei. 
The wave functions 
of the nucleons with positive energy are used to calculate the contributions
of the valence nucleons. In principle, one should also use the wave functions
of the anti-nucleons to evaluate the contributions of the bound states
emerging from the Dirac sea. Considering
the difficulties caused by the large 
number of anti-nucleons in bound states in any practical numerical
procedure, we will employ the technique of
derivative expansion developed in Ref. \cite{Ait84} to compute the
 effects of the Dirac sea, including the contributions of the one-nucleon loop
as well as the one-meson loop. The paper is organized as follows: In Sect. II
we introduce the effective Lagrangian used here. In Sect. III 
the plane-wave solutions of
the Dirac equation in homogeneous nuclear matter are discussed. 
In Sect. IV we constitute
the RHA approach for the bound states of positive energy and negative energy. 
In Sect. V we present the numerical results and discussions.
Finally, a summary and outlook is given in Sect. VI. 

 \end{sloppypar}
\begin{center}
{\bf II. EFFECTIVE LAGRANGIAN}
\end{center}
\begin{sloppypar}
We start from the Lagrangian density for nucleons interacting through the 
exchange of mesons \cite{Ser86,Rei86,Gam90}
\begin{equation}
{\cal L}={\cal L}_{\rm F}+{\cal L}_{\rm I}.
\end{equation}
Here ${\cal L}_{F}$ is the Lagrangian density for free nucleon, mesons 
 and photon 
\begin{eqnarray}
{\cal L}_{\rm F}&=&\bar{\psi}[i\gamma_{\mu}\partial^{\mu}-M_{N}]\psi
   + \frac{1}{2}
\partial_{\mu}\sigma\partial^{\mu}\sigma-U(\sigma)   
 -\frac{1}{4}\omega_{\mu\nu}\omega^{\mu\nu} \nonumber \\
&& + \frac{1}{2}m_{\omega}^{2}\omega_{\mu}\omega^{\mu}            
 - \frac{1}{4} {\bf R}_{\mu\nu}{\bf R}^{\mu\nu}
 +\frac{1}{2}m_{\rho}^{2}{\bf R}_{\mu} \cdot {\bf R}^{\mu}  
- \frac{1}{4}  A_{\mu\nu} A^{\mu\nu}
    \end{eqnarray}
and U($\sigma$) is the self-interaction part of the scalar field
\cite{Bog83} 
\begin{equation}
  U(\sigma)=
   \frac{1}{2}m_{\sigma}^{2}\sigma^{2}+\frac{1}{3!}b
\sigma^{3}+\frac{1}{4!}c\sigma^{4}.
\end{equation}
In the above expressions $\psi$ is the Dirac spinor of the nucleon;
$\sigma$, $\omega_{\mu}$, ${\bf R}_{\mu}$ and $A_{\mu}$ represent the scalar
meson, vector meson, isovector-vector meson field and the electromagnetic
field, respectively. Here the field tensors for the omega, rho and photon
are given in terms of their potentials by
 \begin{eqnarray}
&&  \omega_{\mu\nu}=\partial_{\mu}\omega_{\nu}
   -\partial_{\nu}\omega_{\mu},\\
&&  {\bf R}_{\mu\nu}=\partial_{\mu}{\bf R}_{\nu} -\partial_{\nu}{\bf
R}_{\mu}, \\
&&   A_{\mu\nu}=\partial_{\mu}A_{\nu} -\partial_{\nu}A_{\mu}.
  \end{eqnarray}
${\cal L}_{I}$ is the interaction Lagrangian density
\begin{equation}
 {\cal L}_{I}={\rm g}_{\sigma}\bar{\psi}\psi\sigma
      - {\rm g}_{\omega}\bar{\psi}\gamma_{\mu}\psi\omega^{\mu} 
 - \frac{1}{2}{\rm g}_{\rho}\bar{\psi}\gamma_{\mu}\mbox{\boldmath $\tau$}
\cdot \psi {\bf R}^{\mu} - \frac{1}{2} e \bar{\psi}(1+\tau_{0})\gamma_{\mu}
 \psi A^{\mu}.
\end{equation} 
Here $\mbox{\boldmath $\tau$}$ is the isospin operator of the
 nucleon and $\tau_{0}$ is its third component. ${\rm g}_{\sigma}$, ${\rm
g}_{\omega}$, ${\rm g}_{\rho}$ and $e^{2}/4\pi = 1/137$ are the coupling
strengths for the $\sigma$-, $\omega$-, $\rho$-meson  and  for the photon,
respectively. $M_{N}$ is the free nucleon mass and $m_{\sigma}$,
$m_{\omega}$, $m_{\rho}$ are the masses of the $\sigma$-, $\omega$-, 
 and $\rho$-meson.
For simplicity, in the following discussions we consider the
$\sigma$- and $\omega$-exchange explicitly. The relevant formulae for the
$\rho$-exchange and the electromagnetic field can be obtained in a 
straightforward way after calculating the $\omega$-exchange. The detailed
expressions for all these meson fields and the electromagnetic field 
 will be given in Sect. IV.

\end{sloppypar}
\begin{center}
{\bf III. PLANE-WAVE SOLUTIONS OF THE DIRAC EQUATION IN STATIC NUCLEAR MATTER}
\end{center}
\begin{sloppypar}
From the effective Lagrangian given in the above section, we obtain the 
Dirac equation
 \begin{equation}
 \left[ i\gamma_{\mu}\partial^{\mu} - {\rm g}_{\omega}\gamma_{\mu}\omega^{\mu}
 -M_{N} + {\rm g}_{\sigma}\sigma \right] \psi =0 .
 \end{equation}
In static nuclear matter, this reads
 \begin{equation}
 i \frac{\partial}{\partial t}\psi = \left[ -i \mbox{\boldmath $\alpha$} \cdot
 \mbox{\boldmath $\nabla          $} + {\rm g}_{\omega}\omega_{0} + \beta
 (M_{N} - {\rm g}_{\sigma}\sigma) \right] \psi.
 \end{equation}
We know that Eq. (10) has two solutions, i.e., the positive-energy solution
$E_{+}$ and the negative-energy solution $E_{-}$, which satisfy the following
equations 
 \begin{eqnarray}
&&  \left( E_{+} - {\rm g}_{\omega}\omega_{0} \right) {\cal U} ({\bf p})
 =\left[ \mbox{\boldmath $\alpha$} \cdot {\bf p} + \beta \left(  M_{N}
 - {\rm g}_{\sigma}\sigma \right) \right] {\cal U} ({\bf p}), \\
&&  \left( E_{-} - {\rm g}_{\omega}\omega_{0} \right) {\cal V}(-{\bf p})
 =\left[ \mbox{\boldmath $\alpha$} \cdot {\bf p} + \beta \left( M_{N}
 - {\rm g}_{\sigma}\sigma \right) \right] {\cal V} (-{\bf p}).   
 \end{eqnarray} 
Here ${\cal U}({\bf p})$ and ${\cal V}({\bf p})$ are two spinors \cite{Gre89}
  \begin{equation}
 {\cal U}({\bf p}) = N \left( \begin{array}{c}
 \chi_{\sigma} \\ \frac{\mbox{\boldmath $\sigma$}\cdot{\bf p}}{E^{*}({\bf p})
 + m^{*}} \chi_{\sigma} \end{array} \right) \;\;\;\;\;\;\;\;\;\;\;\;\; 
 {\cal V}({\bf p}) = N \left( \begin{array}{c}
  \frac{\mbox{\boldmath $\sigma$}\cdot{\bf p}}{E^{*}({\bf p})
 + m^{*}} \chi_{\sigma} \\ \chi_{\sigma} \end{array} \right), 
  \end{equation}
where $N$ is the normalization factor and 
  \begin{eqnarray}
&& m^{*}=M_{N}-{\rm g}_{\sigma}\sigma, \\
&& E^{*}({\bf p})=\sqrt{{\bf p}^{2} + m^{*2}}.
  \end{eqnarray}
Defining the effective positive-energy $E_{+}^{*}$ and the effective 
negative-energy $E_{-}^{*}$ as
 \begin{equation}
 E_{+}^{*}=E_{+} - {\rm g}_{\omega}\omega_{0} \;\;\;\;\;\;\;\;\;\;\;\;\;\;\;\;
 E_{-}^{*}=E_{-} - {\rm g}_{\omega}\omega_{0},
 \end{equation}
together with Eqs. (11) $\sim$ (13) one can perform calculations 
similar to the free Dirac equation and obtains
  \begin{equation}
 E^{*}_{+}=E^{*}({\bf p}) \;\;\;\;\;\;\;\;\;\;\;\;\;\;
 E^{*}_{-}=-E^{*}({\bf p}).
  \end{equation}
Therefore,
  \begin{eqnarray}
&& E_{+}=\sqrt{{\bf p}^{2} + m^{*2}} + {\rm g}_{\omega}\omega_{0}, \\
&& E_{-}=- \sqrt{{\bf p}^{2} + m^{*2}} + {\rm g}_{\omega}\omega_{0}.    
  \end{eqnarray}
The wave packet $\psi$ can be expanded according to the plane waves of
positive energy and negative energy as
 \begin{equation}
\psi({\bf x},t)=\int\, \frac{d^{3}p}{(2\pi)^{3/2}} \left( \frac{m^{*}}{E^{*}
 ({\bf p})} \right)^{1/2}\sum_{\pm s} \left[ b_{{\bf p},s}{\cal U}({\bf p},s)
 e ^{i {\bf p}\cdot {\bf x} - iE_{+}t }
 + d^{+}_{{\bf p},s}{\cal V}(-{\bf p},s)
 e ^{i {\bf p}\cdot {\bf x} - iE_{-}t } \right].
 \end{equation}
The coefficient $b_{{\bf p},s}$ are the probability amplitudes for waves with
positive energy, whereas $d^{+}_{{\bf p},s}$ are those for negative energy.
Both the nucleons with positive energy and negative energy transport forward
in time.
According to the Dirac's hole theory, each particle has its partner 
anti-particle which transports backward in time.  
One can write down the corresponding equations for the 
anti-nucleons which have the opposite signs for the energy and momentum 
term compared to the eigenequations of the nucleons                         
 \begin{eqnarray}
&& - \left( \bar{E}_{+} + {\rm g}_{\omega}\omega_{0} \right) {\cal V}({\bf p})
 =\left[ - \mbox{\boldmath $\alpha$} \cdot {\bf p} + \beta \left( M_{N}
 - {\rm g}_{\sigma}\sigma \right) \right] {\cal V} ({\bf p}), \\
&& - \left( \bar{E}_{-} + {\rm g}_{\omega}\omega_{0} \right) {\cal U}(-{\bf p})
 =\left[- \mbox{\boldmath $\alpha$} \cdot {\bf p} + \beta \left( M_{N}
 - {\rm g}_{\sigma}\sigma \right) \right] {\cal U} (-{\bf p}).   
 \end{eqnarray} 
Here $\bar{E}_{+}$, $\bar{E}_{-}$ are the positive energy and negative
energy of the anti-nucleon and ${\cal V}({\bf p})$, ${\cal U}(-{\bf p})$
are their corresponding eigenfunctions, respectively.
Again, we define the effective positive-energy $\bar{E}_{+}^{*}$ and 
the effective negative-energy $\bar{E}_{-}^{*}$ of the anti-nucleon as 
 \begin{equation}
 \bar{E}_{+}^{*}=\bar{E}_{+}+ {\rm g}_{\omega}\omega_{0} 
 \;\;\;\;\;\;\;\;\;\;\;\;\;\;\;\;
 \bar{E}_{-}^{*}=\bar{E}_{-} + {\rm g}_{\omega}\omega_{0}.
 \end{equation}
With the spinors of Eq. (13) we obtain
  \begin{equation}
 \bar{E}^{*}_{+}=E^{*}({\bf p}) \;\;\;\;\;\;\;\;\;\;\;\;\;\;
 \bar{E}^{*}_{-}=-E^{*}({\bf p}) 
  \end{equation}
and
  \begin{eqnarray}
&& \bar{E}_{+}=\sqrt{{\bf p}^{2} + m^{*2}} - {\rm g}_{\omega}\omega_{0}, \\
&& \bar{E}_{-}=- \sqrt{{\bf p}^{2} + m^{*2}} - {\rm g}_{\omega}\omega_{0}.    
  \end{eqnarray}
From Eqs. (18), (19) and (25) , (26) one can easily find the relations
  \begin{equation}
 \bar{E}_{+}=-E_{-} \;\;\;\;\;\;\;\;\;\;\;\;\;\;
 \bar{E}_{-}=-E_{+}.
  \end{equation}
Thus, in order to obtain the full spectrum of the Dirac equation, one can solve
the equations for the nucleon with positive energy and negative energy, i.e., 
Eqs. (11) and (12). Alternatively, one can solve the Dirac equations for the
nucleon and the anti-nucleon but both of them with positive energy, 
 i.e., Eqs. (11)
and (21). The wave packet $\psi$ can also be expanded according to the plane
waves of the nucleon and the anti-nucleon as
 \begin{equation}
\psi({\bf x},t)=\int\, \frac{d^{3}p}{(2\pi)^{3/2}} \left( \frac{m^{*}}{E^{*}
 ({\bf p})} \right)^{1/2}\sum_{\pm s} \left[ b_{{\bf p},s}{\cal U}({\bf p},s)
 e ^{i {\bf p}\cdot {\bf x} - iE_{+}t }
 + d^{+}_{{\bf p},s}{\cal V}({\bf p},s)
 e ^{-i {\bf p}\cdot {\bf x} + i\bar{E}_{+}t } \right].
 \end{equation}
Here now both the nucleon as well as
 the anti-nucleon have positive energy. The nucleon
transports forward in time 
while the anti-nucleon transports backward in time which can be seen from 
the sign of time in exponential functions. In relativistic 
quantum field theory, $b^{+}_{{\bf p},s}$ and $d^{+}_{{\bf p},s}$ 
are explained as 
the creation operators for the nucleon and the anti-nucleon, respectively. This
is the main strategy which will be used in our following considerations for
finite nuclei. Instead of searching for two solutions with positive energy and
negative energy for the Dirac equation of the nucleon, 
we solve the Dirac equations
for the nucleon and the anti-nucleon simultaneously. For each equation we 
look for only one solution, that is, the positive-energy solution.

When the negative energy of the nucleon $E_{-}$ is larger than the nucleon 
free mass, the system becomes unstable with respect to the nucleon--anti-nucleon
 pair creation. At zero-momentum, one has the critical condition
 \begin{equation}
 {\rm g}_{\sigma}\sigma + {\rm g}_{\omega}\omega_{0}=2M_{N},
 \end{equation}
that is,
 \begin{equation}
 V-S=2M_{N}
 \end{equation}
if one defines
  \begin{equation}
 S=-{\rm g}_{\sigma}\sigma \;\;\;\;\;\;\;\;\;\;\;\;\;\;
 V={\rm g}_{\omega}\omega_{0}.
  \end{equation}
\end{sloppypar}
\begin{center}
{\bf IV. RELATIVISTIC HARTREE APPROXIMATION OF FINITE NUCLEI}
\end{center}
\begin{sloppypar}
In finite nuclei the meson fields in Eq. (10) are space-dependent. The
field operator of the Dirac equation can be written as 
  \begin{equation}
\psi({\bf x},t)=\sum_{\alpha} \left[ b_{\alpha}\psi_{\alpha}({\bf x})
 e^{-i E_{\alpha}t} + d^{+}_{\alpha}\psi^{a}_{\alpha}({\bf x}) e^{i \bar{E}
 _{\alpha} t} \right].
  \end{equation}
Here the label $\alpha$ denotes the full set of single-particle quantum
numbers. $\psi_{\alpha}({\bf x})$ are the wave functions of nucleons and
$\psi_{\alpha}^{a}({\bf x})$ are those of anti-nucleons; $E_{\alpha}$ and
$\bar{E}_{\alpha}$ are their positive energies, respectively. $b^{+}_{\alpha}$
and $d^{+}_{\alpha}$ are nucleon and anti-nucleon creation operators that
satisfy the standard anticommutation relations. We assume that the meson fields
depend only on the radius and discuss the problem in spherically symmetric
nuclei. In this case, the usual angular momentum and parity are good quantum
numbers. As described in Refs. \cite{Gre89,Bjo64}, eigenfunctions of the 
angular momentum and the parity operator are the well-known spherical spinors.
We make the following ansatz for the wave functions of nucleons
  \begin{equation}
 \psi_{\alpha}({\bf x})= \left( \begin{array}{l}
 i \frac{G_{\alpha}(r)}{r} \Omega_{jlm}(\frac{{\bf r}}{r})  \\
  \frac{F_{\alpha}(r)}{r}\frac{\mbox{\boldmath $\sigma$}\cdot {\bf r}}{r}
 \Omega_{jlm}(\frac{{\bf r}}{r}) \end{array} \right)
  \end{equation}
and anti-nucleons
  \begin{equation}
 \psi_{\alpha}^{a}({\bf x})= \left( \begin{array}{l}
 \frac{\bar{F}_{\alpha}(r)}{r}\frac{\mbox{\boldmath $\sigma$}\cdot {\bf r}}{r}
 \Omega_{jlm}(\frac{{\bf r}}{r})  \\      
  i \frac{\bar{G}_{\alpha}(r)}{r} \Omega_{jlm}(\frac{{\bf r}}{r}) 
 \end{array} \right).
  \end{equation}
Here $\Omega_{jlm}$ are the spherical spinors defined as \cite{Gre89}
  \begin{equation}
\Omega_{jlm}=\sum_{m^{\prime}m_{s}}\left( l \frac{1}{2} j \mid m^{\prime}
 m_{s} m \right) Y_{l m^{\prime}} \chi_{\frac{1}{2}m_{s}},
  \end{equation}
$Y_{l m^{\prime}}$ are the spherical harmonics and $\chi_{\frac{1}{2}m_{s}}$
are the eigenfunctions of the spin operators. $G_{\alpha}$, $F_{\alpha}$ and
 $\bar{F}_{\alpha}$, $\bar{G}_{\alpha}$ are the remaining {\em real} radial
wave functions of nucleons and anti-nucleons for upper and lower components,
respectively. Applying the parity operator $\hat{P}=e^{i \phi}\beta \hat{P}
 _{0}$, $\hat{P}_{0}$ changes ${\bf x}$ into $ - {\bf x}$, to Eqs. (33)
and (34), one can easily find that the $\psi_{\alpha}({\bf x})$ has the 
opposite eigenvalue of parity, $(-1)^{l}$, to the $\psi_{\alpha}^{a}({\bf x})$,
$(-1)^{l+1}$, as it should be \cite{Gre89}.

Inserting Eqs. (33) and (34) into Eq. (10) ($\sigma \rightarrow \sigma(r)$
and $\omega_{0} \rightarrow \omega_{0}(r)$), we obtain the coupled  
radial wave functions for the nucleon
  \begin{eqnarray}
&& E_{\alpha}G_{\alpha}(r)=\left( - \frac{d}{dr} + \frac{k_{\alpha}}{r} \right)
 F_{\alpha}(r) + \left( M_{N} - {\rm g}_{\sigma}\sigma(r) + {\rm g}_{\omega}
 \omega_{0}(r) \right) G_{\alpha}(r), \\
&& E_{\alpha}F_{\alpha}(r)=\left( \frac{d}{dr} + \frac{k_{\alpha}}{r} \right)
 G_{\alpha}(r) + \left( - M_{N} + {\rm g}_{\sigma}\sigma(r) + {\rm g}_{\omega}
 \omega_{0}(r) \right) F_{\alpha}(r)    
  \end{eqnarray}
and the anti-nucleon
  \begin{eqnarray}
&& -\bar{E}_{\alpha}\bar{F}_{\alpha}(r)=\left(\frac{d}{dr} 
+ \frac{k_{\alpha}}{r} \right)
 \bar{G}_{\alpha}(r) + \left( M_{N} - {\rm g}_{\sigma}\sigma(r) 
+ {\rm g}_{\omega}
 \omega_{0}(r) \right) \bar{F}_{\alpha}(r), \\
&& -\bar{E}_{\alpha}\bar{G}_{\alpha}(r)=\left( - \frac{d}{dr} 
+ \frac{k_{\alpha}}{r} \right)
 \bar{F}_{\alpha}(r) + \left( -M_{N} + {\rm g}_{\sigma}\sigma(r) 
+ {\rm g}_{\omega}
 \omega_{0}(r) \right) \bar{G}_{\alpha}(r)    
  \end{eqnarray}
respectively, where
  \begin{equation}
k_{\alpha}= \left\{  \begin{array}{ll} 
-(l+1) & \mbox{for $j=l+\frac{1}{2}$} \\
 l & \mbox{for $j=l-\frac{1}{2}$} \end{array} \right.   .
  \end{equation}
In order to resemble Schr\"{o}dinger equations for the Dirac equations, 
we eliminate the small 
components. For the nucleon we eliminate the lower
component while for the anti-nucleon the upper component. By defining
the Schr\"{o}dinger equivalent effective mass and potential of  the nucleon
 \begin{eqnarray}
&&  M_{eff}= E_{\alpha} - {\rm g}_{\omega}\omega_{0}(r) + M_{N} - 
 {\rm g}_{\sigma}\sigma (r), \\
&& U_{eff}= M_{N} - {\rm g}_{\sigma}\sigma (r)
 + {\rm g}_{\omega}\omega_{0} (r)
 \end{eqnarray}
and the anti-nucleon
 \begin{eqnarray}
&&  \bar{M}_{eff}= \bar{E}_{\alpha} + {\rm g}_{\omega}\omega_{0}(r) + M_{N} - 
 {\rm g}_{\sigma}\sigma (r), \\
&& \bar{U}_{eff}= M_{N} - {\rm g}_{\sigma}\sigma (r)
 - {\rm g}_{\omega}\omega_{0} (r),
 \end{eqnarray}
we arrive at the Schr\"{o}dinger equations for the upper component 
of the nucleon's  wave function
 \begin{equation}
E_{\alpha}G_{\alpha}(r) = \left(- \frac{d}{dr} + \frac{k_{\alpha}}{r}
 \right) M_{eff}^{-1} \left( \frac{d}{dr} + \frac{k_{\alpha}}{r} \right)
 G_{\alpha}(r) + U_{eff}G_{\alpha}(r)
 \end{equation}
and the lower component of the anti-nucleon's wave function
 \begin{equation}
\bar{E}_{\alpha}\bar{G}_{\alpha}(r) = \left(- \frac{d}{dr}+\frac{k_{\alpha}}{r}
 \right) \bar{M}_{eff}^{-1} \left( \frac{d}{dr} + \frac{k_{\alpha}}{r} \right)
 \bar{G}_{\alpha}(r) + \bar{U}_{eff}\bar{G}_{\alpha}(r).
 \end{equation}
The small components can be obtained through the following relations
 \begin{eqnarray}
&& F_{\alpha}(r) = M_{eff}^{-1} \left( \frac{d}{dr} + \frac{k_{\alpha}}{r}
 \right) G_{\alpha}(r), \\
&& \bar{F}_{\alpha}(r) = - \bar{M}_{eff}^{-1} 
\left( \frac{d}{dr} + \frac{k_{\alpha}}{r}
 \right) \bar{G}_{\alpha}(r).     
 \end{eqnarray}
Of course, the radial wave functions are normalized
 \begin{eqnarray}
 && \int^{\infty}_{0} dr \left[ \mid G_{\alpha}(r) \mid ^{2}
 + \mid F_{\alpha} (r) \mid ^{2} \right] =1 , \\
 && \int^{\infty}_{0} dr \left[ \mid \bar{G}_{\alpha}(r) \mid ^{2}
 + \mid \bar{F}_{\alpha} (r) \mid ^{2} \right] =1 .   
 \end{eqnarray}
From Eqs. (45) and (46) one finds that the Schr\"{o}dinger equation of
the anti-nucleon has the same form as that of the nucleon. The only difference
relies on the definition of the effective mass and potential, that is, 
the vector 
field changes its sign. The so-called {\em G-parity} comes out automatically.
Eqs. (45) and (46) can be solved numerically by the standard technique as 
described in Ref. \cite{Rei86}. The single-particle energy of the nucleon 
and the anti-nucleon can be written as
 \begin{eqnarray}
 E_{\alpha} &=& \int^{\infty}_{0} dr \lbrace G_{\alpha}(r) \left( 
 -\frac{d}{dr} + \frac{k_{\alpha}}{r} \right) F_{\alpha}(r) + F_{\alpha}(r)
 \left( \frac{d}{dr} + \frac{k_{\alpha}}{r} \right) G_{\alpha}(r) 
 \nonumber \\
&&   + G_{\alpha}(r)U_{eff}G_{\alpha}(r) - F_{\alpha}(r) \left( M_{eff}
 - E_{\alpha} \right) F_{\alpha}(r) \rbrace, \\ 
 \bar{E}_{\alpha} &=& \int^{\infty}_{0} dr \lbrace - \bar{G}_{\alpha}(r) \left( 
 -\frac{d}{dr} + \frac{k_{\alpha}}{r} \right)\bar{F}_{\alpha}(r) 
 - \bar{F}_{\alpha}(r)
 \left( \frac{d}{dr} + \frac{k_{\alpha}}{r} \right) \bar{G}_{\alpha}(r) 
  \nonumber \\
&& + \bar{G}_{\alpha}(r)\bar{U}_{eff}\bar{G}_{\alpha}(r) 
 - \bar{F}_{\alpha}(r) \left( \bar{M}_{eff}
 - \bar{E}_{\alpha} \right) \bar{F}_{\alpha}(r) \rbrace,  
 \end{eqnarray}
which are obtained through the iteration procedure. The negative energies of 
nucleons are just the minus sign of $\bar{E}_{\alpha}$.

The meson fields in Eqs. (41 ) $\sim$ (44) are determined by the Laplace 
equations
 \begin{eqnarray}
&& \left( \mbox{\boldmath $\nabla          $}^{2} - m_{\sigma}^{2} \right)
 \sigma(r) = - {\rm g}_{\sigma} \rho_{S}(r) + \frac{1}{2}b\sigma^{2}
 + \frac{1}{3!}c \sigma^{3}, \\
&& \left( \mbox{\boldmath $\nabla          $}^{2} - m_{\omega}^{2} \right)
 \omega_{0} (r) = - {\rm g}_{\omega} \rho_{0} (r),
 \end{eqnarray}
where 
 \begin{eqnarray}
&& \rho_{S}(r) = \sum_{i= - \infty}^{+ \infty} w_{i}
   \bar{\psi}_{i}\psi_{i} + CT, \\
&& \rho_{0}(r) = \sum_{i= - \infty}^{+ \infty} w_{i}
   \psi^{+}_{i}\psi_{i} + CT,   
 \end{eqnarray}
here $w_{i}$ are the occupation numbers defined by the 
 creation and annihilation operators of Eq. (32) as usual \cite{Rei86}. 
The $CT$ are the counterterms. The sums in Eqs. (55) and (56)
run over the full spectrum of the Dirac equation: the negative-energy 
continuum (the positive-energy continuum of the anti-nucleon),  some
negative-energy bound states (positive-energy bound states of the anti-nucleon),
 the positive-energy bound states which correspond to the usual nucleon shell
model states and the positive-energy continuum (included through the 
renormalization procedure). 

Let us first assume that
the positive-energy continuum and the negative-energy continuum together with
the contributions of the counter terms will yield finite terms for
$\Delta \rho_{S}(r)$  and $\Delta \rho_{0}(r)$. 
 The scalar and baryon density can then be expressed as
 \begin{eqnarray}
 \rho_{S}(r)&=&\frac{1}{4\pi r^{2}}\sum_{\alpha} w_{\alpha}(2 j_{\alpha}
 + 1)(G_{\alpha}^{2}(r) - F_{\alpha}^{2} (r)) \nonumber \\
&& + \frac{1}{4\pi r^{2}}\sum_{\beta} w_{\beta}(2 j_{\beta}
+ 1)(\bar{G}_{\beta}^{2}(r) - \bar{F}_{\beta}^{2} (r)) + \Delta \rho_{S}(r), \\
 \rho_{0}(r)&=&\frac{1}{4\pi r^{2}}\sum_{\alpha} w_{\alpha}(2 j_{\alpha}
 + 1)(G_{\alpha}^{2}(r) + F_{\alpha}^{2} (r)) \nonumber \\
&& - \frac{1}{4\pi r^{2}}\sum_{\beta} w_{\beta}(2 j_{\beta}
+ 1)(\bar{G}_{\beta}^{2}(r) + \bar{F}_{\beta}^{2} (r)) - \Delta \rho_{0}(r),   
 \end{eqnarray}
The first terms
on the right-hand-side of Eqs. (57) and (58) denote the contributions of the
shell model states while the second terms represent the contributions of 
the bound states of negative energy. If one thinks about that there might exist
twenty thousand nucleons in the bound states of negative energy, 
it is obviously not 
practical to compute the contributions of those bound states through 
evaluating their wave functions. Fortunately, a rather elegant technique has
been developed by several authors \cite{Ait84}, which  takes into account the 
vacuum contributions to the source term of the meson fields in finite nuclei. 
Let us write 
 \begin{eqnarray}
&& \rho_{S}(r)=\rho_{S}^{val}(r)+\rho_{S}^{sea}(r), \\
&& \rho_{0}(r)=\rho_{0}^{val}(r)+\rho_{0}^{sea}(r),   
 \end{eqnarray}
where $\rho_{S}^{val}(r)$ and $\rho_{0}^{val}(r)$ are just the first terms  
on the right-hand-side of Eqs. (57) and (58). $\rho_{S}^{sea}(r)$ and $\rho_{0}
^{sea}(r)$ are the contributions of the vacuum, including the bound states and
the continuum as well as the counterterm contributions. Originally, vacuum
corrections for finite nuclei (in one-loop approximation) 
were included only in a local density 
approximation \cite{Ser86}. Later on, Perry \cite{Per86} and Wasson \cite{Was88}
 considered derivative corrections to the nuclear matter results. It was found
that the leading derivative correction is of the same order of magnitude as
the effective potential (i.e., the nuclear matter results) while the next-order
derivative correction is two or three orders of
 magnitude smaller than the leading order.
That means that the derivative expansion converges rapidly.
Since we include the non-linear self-interaction of the 
scalar field in the model, the
contributions of the one-meson loop from the scalar meson should be  taken into
account in addition to the one-nucleon loop. The contributions of the one-meson
loop have been calculated in Ref. \cite{Fox89} 
up to the effective potential term.
Here this is extended  to include the leading derivative correction. 

The effective action of the system at the one-loop level can be written as
  \begin{equation}
\Gamma = \int\, d^{4}x \left(\frac{1}{2} 
\partial_{\mu}\sigma \partial^{\mu}\sigma
 -U(\sigma) - \frac{1}{4}\omega_{\mu\nu}\omega^{\mu\nu} + \frac{1}{2}m_{\omega}
^{2}\omega_{\mu}\omega^{\mu} + CT\right) 
+ \Gamma_{{\rm valence}}+\Gamma^{(1)}(\sigma)
 + \Gamma^{(1)}(\psi). 
  \end{equation} 
Here $\Gamma_{{\rm valence}}$ is the contribution from the valence nucleons, 
which for time independent background fields is just minus the energy of the
valence nucleons. $\Gamma^{(1)}(\sigma)$ and $\Gamma^{(1)}(\psi)$ represent 
the contributions of the Dirac sea stemming from the one-meson loop and 
one-nucleon loop, respectively. They are defined as \cite{Jac74}
  \begin{eqnarray}
 && \Gamma^{(1)}(\sigma)=\frac{i}{2} \hbar {\rm Tr}\ln \left[ P^{2} -
 (m_{\sigma}^{2} + b\sigma + \frac{1}{2} c\sigma^{2} ) \right], \\
 && \Gamma^{(1)}(\psi) = -i \hbar {\rm Tr} \ln \left[ \not\!\!P - m^{*}
 - {\rm g}_{\omega}\not\!\!\omega \right],
  \end{eqnarray}
where the trace is over spatial and internal variables. $\Gamma^{(1)}(\sigma)$
and $\Gamma^{(1)}(\psi)$ can be expanded in powers of the derivatives of
the scalar and vector fields. Using Lorentz invariance one can determine the
functional Taylor series as
 \begin{eqnarray}
&& \Gamma^{(1)}(\sigma)=\int\, d^{4}x \left( -V_{B}^{(1)}(\sigma) + \frac{1}{2}
 Z^{(1)}(\sigma)(\partial_{\mu} \sigma)^{2} + Y^{(1)}(\sigma)
 (\partial_{\mu}\sigma)
 ^{4} + ... \right), \\
&& \Gamma^{(1)}(\psi)=\int\, d^{4}x \left( - V_{F}^{(1)}(\sigma) + \frac{1}{2}
 Z_{1\sigma}^{(1)}(\sigma)(\partial_{\mu}\sigma)^{2} + \frac{1}{4}Z^{(1)}
 _{1\omega}(\sigma) \omega_{\mu\nu}\omega^{\mu\nu} + ... \right).
 \end{eqnarray}
Here $V_{B}^{(1)}(\sigma)$ and $V_{F}^{(1)}(\sigma)$ are the effective 
potentials from the one-boson loop and one-fermion loop, in which the field
is a constant, $\sigma(x)=\sigma_{0}$, the same situation as in nuclear matter.
These two terms contain divergent part and should be renormalized. Through
adding the suitable counterterms, $V_{B}^{(1)}(\sigma)$ and $V_{F}^{(1)}
 (\sigma)$ can be evaluated in nuclear matter which turn out to be
\cite{Chi77,Ser86}
  \begin{eqnarray}
V_{B}^{(1)}(\sigma)&=& \frac{m_{\sigma}^{4}}{(8\pi)^{2}} \left[ \left( 1 + 
 \frac{b\sigma}{m_{\sigma}^{2}} + \frac{c\sigma^{2}}{2m_{\sigma}^{2}} \right)
 ^{2} \ln \left( 1+ \frac{b\sigma}{m_{\sigma}^{2}} + \frac{c\sigma^{2}}
{2m_{\sigma}^{2}} \right) - \left( \frac{b\sigma}{m_{\sigma}^{2}} + \frac{
 c\sigma^{2}}{2m_{\sigma}^{2}} \right) \right. \nonumber \\
 && \left. 
 -\frac{3}{2} \left( \frac{b\sigma}{m_{\sigma}^{2}} + \frac{c\sigma^{2}}
 {2m_{\sigma}^{2}} \right)^{2} - \frac{1}{3} \left( \frac{b\sigma}{m_{\sigma}
 ^{2}} \right) ^{2} \left( \frac{b\sigma}{m_{\sigma}^{2}} + \frac{3c\sigma^{2}}
 {2m_{\sigma}^{2}} \right) + \frac{1}{12} \left( \frac{b\sigma}{m_{\sigma}^{2}}
 \right) ^{4} \right], \\
V_{F}^{(1)}(\sigma)&=& - \frac{1}{4\pi^{2}} \left[ \left( M_{N} - 
{\rm g}_{\sigma}\sigma \right)^{4}\ln \left( 1- \frac{{\rm g}_{\sigma}\sigma}
{M_{N}} \right) + M_{N}^{3}{\rm g}_{\sigma}\sigma - \frac{7}{2}M_{N}^{2}
 {\rm g}_{\sigma}^{2}\sigma^{2} \right. \nonumber \\
 && \left. + \frac{13}{3}M_{N}{\rm g}_{\sigma}^{3}
 \sigma^{3} - \frac{25}{12}{\rm g}_{\sigma}^{4}\sigma^{4} \right].
  \end{eqnarray}
As discussed above, the derivative expansion converges rapidly. Thus, here 
we consider only the lowest order derivative terms in Eqs. (64) and (65).
By means of the technique developed in Ref. \cite{Ait84}, one can determine the
functional coefficients of the leading derivative correction, which read
  \begin{eqnarray}
&&Z^{(1)}(\sigma)=\frac{1}{12}\frac{(b+c\sigma)^{2}}
 {16\pi^{2}(m_{\sigma}^{2} + b\sigma
 +\frac{1}{2}c\sigma^{2}) }, \\
&& Z^{(1)}_{1\sigma}(\sigma)= - \frac{{\rm g}_{\sigma}^{2}}{2\pi^{2}} \ln
 \left( \frac{m^{*}}{M_{N}} \right), \\
&& Z^{(1)}_{1\omega}(\sigma)=  \frac{{\rm g}_{\omega}^{2}}{3\pi^{2}} \ln
 \left( \frac{m^{*}}{M_{N}} \right). 
  \end{eqnarray}
The above expressions are finite, and independent of any renormalization 
conditions. The divergent integral appearing in $\Gamma^{(1)}(\sigma)$ and
$\Gamma^{(1)}(\psi)$ are included in the effective potential terms of
$V_{B}^{(1)}(\sigma)$ and $V_{F}^{(1)}(\sigma)$. The equations of meson fields,
i.e., Eqs. (53) and (54), can be obtained through minimizing the effective 
action of Eq. (61) with respect to the corresponding fields. 
 With the definitions
 of Eqs. (57) $\sim$ (60), $\Gamma_{{\rm valence}}$ contributes to the $\rho
^{val}_{S}(r)$ and $\rho_{0}^{val}(r)$ while $\Gamma^{(1)}(\sigma)$ and
 $\Gamma^{(1)}(\psi)$ contribute to the $\rho^{sea}_{S}(r)$ and $\rho^{sea}
 _{0}(r)$.
Inserting Eqs. (64) $\sim$ (70) into Eq. (61) and minimizing the effective
action, one obtains 
the concrete expressions of $\rho_{S}^{sea}(r)$ and $\rho_{0}^{sea}(r)$
which read   
 \begin{eqnarray}
\rho_{S}^{sea}(r) &=& 
 -\frac{1}{{\rm g}_{\sigma}} \frac{\partial}{\partial \sigma}
 \left[V_{B}^{(1)}(\sigma) + V_{F}^{(1)}(\sigma) \right] 
 -\frac{1}{{\rm g}_{\sigma}} \left[ \frac{1}{2} \frac{\partial Z^{(1)}(\sigma)}
 {\partial\sigma} (\mbox{\boldmath $\nabla          $} \sigma)^{2} 
 - \mbox{\boldmath $\nabla          $}\cdot Z^{(1)}(\sigma) \mbox{\boldmath
 $\nabla          $} \sigma \right] \nonumber \\
&& -\frac{{\rm g}_{\sigma}}{2\pi ^{2}}\mbox{\boldmath $\nabla          $}
\cdot \ln \left( \frac{m^{*}}{M_{N}}\right) \mbox{\boldmath $\nabla          $}
\sigma - \frac{{\rm g}_{\sigma}^{2}}{4\pi^{2}m^{*}}( \mbox{\boldmath
$\nabla          $}\sigma)^{2} + \frac{{\rm g}_{\omega}^{2}}{6\pi^{2}m^{*}}
 (\mbox{\boldmath $\nabla          $}\omega_{0})^{2}, \\
\rho_{0}^{sea}(r) &=& - \frac{{\rm g}_{\omega}}{3\pi^{2}} \mbox{\boldmath
 $\nabla          $} \cdot \ln \left(\frac{m^{*}}{M_{N}}\right)\mbox{\boldmath
$\nabla          $}  \omega_{0},
 \end{eqnarray}
where
  \begin{eqnarray}
\frac{\partial V_{B}^{(1)}(\sigma)}{\partial \sigma} &=&
\frac{m_{\sigma}^{4}}{(8\pi)^{2}} \left[ 2\left( 1 + 
 \frac{b\sigma}{m_{\sigma}^{2}} + \frac{c\sigma^{2}}{2m_{\sigma}^{2}} \right)
 \left( \frac{b}{m_{\sigma}^{2}} + \frac{c\sigma}{m_{\sigma}^{2}} \right)
 \ln \left( 1 + \frac{b\sigma}{m_{\sigma}^{2}} + \frac{c\sigma^{2}}{2m_{\sigma}
 ^{2}} \right) \right. \nonumber \\
 && \left. -2 \left( \frac{b\sigma}{m_{\sigma}^{2}} + \frac{c\sigma^{2}}{2m_{\sigma}
 ^{2}} \right) \left( \frac{b}{m_{\sigma}^{2}} + \frac{c\sigma}{m_{\sigma}^{2}}
 \right) - \frac{b^{2}}{m_{\sigma}^{6}} \left( b\sigma^{2} + 2c\sigma^{3}
 \right) + \frac{b^{4}}{3 m_{\sigma}^{8}}\sigma^{3} \right], \\
\frac{\partial V_{F}^{(1)}(\sigma)}{\partial \sigma} &=&
 - \frac{1}{4\pi^{2}} \left[ - {\rm g}_{\sigma} \left( M_{N} - {\rm g}_{\sigma}
 \sigma \right) ^{3} \left( 1 + 4\ln (1-\frac{{\rm g}_{\sigma}\sigma}
 {M_{N}}) \right) + M_{N}^{3}{\rm g}_{\sigma} - 7M_{N}^{2}{\rm g}_{\sigma}^{2}
 \sigma \right. \nonumber \\
 && \left. + 13M_{N}{\rm g}_{\sigma}^{3}\sigma^{2} - \frac{25}{3}{\rm g}
 _{\sigma}^{4}\sigma^{3} \right] 
\end{eqnarray}
and
 \begin{equation}
 \frac{\partial Z^{(1)}(\sigma)}{\partial \sigma}=\frac{1}{192 \pi^{2}}
 \left[ \frac{2c (b+c\sigma)}{(m_{\sigma}^{2}+ b\sigma + \frac{1}{2}c\sigma^{2})
 } - \frac{( b+c\sigma) ^{3}} {(m_{\sigma}^{2} + b\sigma + \frac{1}{2}c \sigma
 ^{2}) ^{2} } \right].
 \end{equation}
Note that $\rho_{0}^{sea}(r)$ is a total derivative and thus the baryon
number is conserved.

To include the contributions of the $\rho$-exchange 
 and the electromagnetic field, 
in Eqs. (41) and (42) we make a replacement of
  \begin{eqnarray}
{\rm g}_{\omega}\omega_{0}(r) \longrightarrow {\rm g}_{\omega}\omega_{0}(r)
 + \frac{1}{2} {\rm g}_{\rho}\tau_{0\alpha}R_{0,0}(r) 
  + \frac{1}{2}e (1+\tau_{0\alpha})A_{0}(r) \nonumber  
  \end{eqnarray}
and in Eqs. (43) and (44)
  \begin{eqnarray}
{\rm g}_{\omega}\omega_{0}(r) \longrightarrow {\rm g}_{\omega}\omega_{0}(r)
 - \frac{1}{2} {\rm g}_{\rho}\tau_{0\alpha}R_{0,0}(r) 
  + \frac{1}{2}e (1+\tau_{0\alpha})A_{0}(r) \nonumber  .
  \end{eqnarray}
Here we have defined that the anti-particle has the same isospin factor as the
corresponding particle. Note that the G-parity of 
$\rho$-meson is positive.
The field equations of the $\rho$-meson and the photon read 
  \begin{eqnarray}
&& ( \mbox{\boldmath $\nabla          $}^{2} - m_{\rho}^{2})R_{0,0}(r)=
 - \frac{1}{2}{\rm g}_{\rho} (\rho_{0,0}^{val}(r) + \rho_{0,0}^{sea}(r) ), \\
&& \mbox{\boldmath $\nabla          $}^{2} A_{0}(r) = -e (\rho_{Pr,0}^{val}
 (r) + \rho_{Pr,0}^{sea}(r) ).
  \end{eqnarray}
We assume isospin-symmetry in the bound states of negative energy, therefore, 
$\rho_{0,0}^{sea}(r)=0$. Other densities appearing in Eqs. (76) and (77) 
can be calculated with the same steps as given above. At the end we obtain
  \begin{eqnarray}
 && \rho_{0,0}^{val}(r)= \frac{1}{4\pi r^{2}} \sum_{\alpha} w_{\alpha} 
 (2j_{\alpha} + 1) \tau_{0\alpha} (G_{\alpha}^{2}(r) + F_{\alpha}^{2}(r) ), \\
 &&\rho_{Pr,0}^{val}(r)=\frac{1}{2}(\rho_{0}^{val}(r) +\rho_{0,0}^{val}(r)), \\
 && \rho_{Pr,0}^{sea}(r)= - \frac{e}{6\pi^{2}} \mbox{\boldmath $\nabla      $}
 \cdot \ln \left(\frac{m^{*}}{M_{N}} \right) \mbox{\boldmath $\nabla          $}
 A_{0}(r).
  \end{eqnarray}
If one includes the $\rho$-exchange and the electromagnetic field in Eq. (63),
a term 
  \begin{displaymath}
\frac{e^{2}}{12\pi ^{2} m^{*}} (\mbox{\boldmath $\nabla          $}A_{0})^{2}
  \end{displaymath}
should be added to Eq. (71), which represents the contribution of the
electromagnetic field to the derivative correction of the one-nucleon loop.
The corresponding contribution of the $\rho$-meson field is negligible 
in the isospin-symmetry vacuum.
The Schr\"{o}dinger equations (45), (46), and the Laplace equations (53), (54),
(76), (77), must be solved numerically in a self-consistent iteration
procedure \cite{Rei86}.

\end{sloppypar}
\begin{center}
{\bf V. NUMERICAL RESULTS AND DISCUSSIONS}
\end{center}
\begin{sloppypar}
Since we employ the derivative expansion to evaluate the contributions 
of the Dirac sea to the source terms of the meson fields,
 the wave functions of anti-nucleons, which are used to calculate the
single-particle energies,  are not involved in evaluating the
vacuum contributions to the scalar and baryon density which are, in turn, 
expressed by means of the scalar and vector field as well as their derivative
terms. The Dirac equation of the nucleon and the equations of
motion of mesons (containing the densities contributed from the vacuum) 
are solved within a self-consistent iteration procedure
\cite{Rei86}. Then, the Dirac equation of the anti-nucleon is solved with the
known mean fields to obtain the wave functions and the single-particle energies 
of anti-nucleons. The space of anti-nucleons are truncated by the specified 
principal and angular quantum numbers $n$ and $j$ with the guarantee  that
the calculated single-particle energies of anti-nucleons are converged when the
truncated space is extended.
We find that the 
results are insensitive to the exact values of $n$ and $j$ provided large 
enough numbers are given. 
We have used $n=4$, $j=9$ for $^{16}{\rm O}$; 
$n=5$, $j=11$ for $^{40}{\rm Ca}$;
and $n=9$, $j=19$ for $^{208}{\rm Pb}$.                       

As pointed out in the Introduction,
in the previous RHA calculations for the bound states of positive energy
\cite{Hor84,Fur89}, the parameters of the model are fitted to the saturation
properties of nuclear matter as well as the $rms$ charge radius in 
$^{40}{\rm Ca}$. The {\em best-fit} routine within the RHA to the properties of 
spherical nuclei has not been performed yet. Thus, we first fit the parameters
of the effective Lagrangian presented in Sect. II 
within the RHA to the empirical data of binding energy,
surface thickness and diffraction radius of 
eight spherical nuclei $^{16}{\rm O}$, $^{40}{\rm Ca}$, 
$^{48}{\rm Ca}$, $^{58}{\rm Ni}$, $^{90}{\rm Zr}$, $^{116}{\rm Sn}$,
$^{124}{\rm Sn}$, and $^{208}{\rm Pb}$
as has been done in
Ref. \cite{Rei86} for the RMF model. 
The experimental values for the observables used in the fit are given
in Table I. In the fitting processes,
we distinguish two different cases with
(RHA1) and without (RHA0) nonlinear self-interaction of the scalar field.
The obtained parameters and the corresponding saturation properties are 
given in Table II. For the sake of comparison, 
two sets of the linear (LIN) and nonlinear (NL1)
RMF parameters from Ref. \cite{Rei86} 
are also presented. One can see that the RHA gives a larger
effective nucleon mass than the RMF does, which is mainly caused by the
feedback of the vacuum to the meson fields, as can be seen from Eqs. (71) $\sim$
(74). When the effective nucleon mass decreases, the
scalar density originated from the Dirac sea $\rho_{S}^{sea}$ increases.  
It is negative
and cancels part of the scalar density contributed from the valence nucleons 
$\rho_{S}^{val}$, which causes the effective nucleon mass to increase again.
At the end, it reaches a balance value. 
In the fitting procedure, we have tried different initial values giving 
smaller effective nucleon mass. After running the code many times, all of them
slowly converge to a large $m^{*}$. 
The larger effective nucleon mass  
 explains why a larger $\chi^{2}$ value is
obtained for the RHA1 compared to the NL1. 
If one uses the current nonlinear RMF/RHA
models to fit the ground states properties of spherical nuclei, an 
effective nucleon mass around 0.6 is preferred. The situation, however, might
be changed when other physical ingredients, e.g., tensor coupling for the
vector fields, correlation effects, 
three-body forces, are taken into account, which warrants further investigation.
On the other hand, in the case of linear model, the RHA0 gives a better fit 
than the LIN does.    
This is mainly due to the vacuum contributions which
improve the theoretical results of the surface thickness substantially, 
and finally
improve the total $\chi^{2}$ value. 

Fig.~1 displays the equations of state of nuclear matter as well as the scalar
and vector potentials as a function of density calculated with four sets of
parameters given in Table II. It is clear that the RHA exhibits a softer equation
of state compared to the RMF, mainly due to the larger $m^{*}$. In the mean 
time, the strengths of the scalar and vector potentials decrease substantially
in the RHA as expected. The same situation happens at finite nuclei. In Fig.~2
and Fig.~3 we present the scalar and vector potentials in $^{16}{\rm O}$, 
$^{40}{\rm Ca}$ and $^{208}{\rm Pb}$ computed with the NL1 and the RHA1 set of
parameters, respectively. The potentials calculated with the RHA1 are about 
half of that calculated with the NL1, implying a strong feedback of the vacuum
to the meson fields. 

As given in Sect. IV, the contribution of the Dirac sea
to the baryon density is a total derivative. The net baryon number is conserved.
Fig.~4 depicts the fractions of the baryon density stemming from the vacuum in 
spherical nuclei of $^{16}{\rm O}$, $^{40}{\rm Ca}$ and $^{208}{\rm Pb}$. It can be
seen that the $\rho_{0}^{sea}$ is more pronounced in light nuclei. In heavy nuclei
it is negligible. Specifically, the $\rho_{0}^{sea} / \rho_{0} \leq 4.0$\% in
$^{16}{\rm O}$, 2.3\% in $^{40}{\rm Ca}$ and 0.6\% in $^{208}{\rm Pb}$.
In Fig.~ 5(a) we compare the baryon densities of $^{16}{\rm O}$ calculated within the 
RHA and the RMF model, respectively. One can see that the difference is not very
siginificant. In the case of the RHA, the contributions of different sources to the
baryon density are shown in Fig. 5(b). The vacuum contribution changes its sign
from the interior to the surface of the nucleus. At large $r$, the Dirac-sea
effect is negligible. This can be observed in Fig.~4 too. For different nuclei,
after the $r$ exceeds the typical values of surface range, the $\rho_{0}^{sea}$
decreases rapidly.

In Fig.~6 the charge densities of three spherical nuclei computed with
the NL1 and the RHA1 set of parameters are compared with the experimental data. 
It seems that 
for light nucleus $^{16}{\rm O}$ the results of the NL1 are closer to the data
than that of the RHA1. Alternatively, for media and heavy nuclei, 
the charge densities calculated with the RHA1 show a better agreement with the data
and less shell fluctuation in the interior of nuclei than that with the NL1 model.
The shell fluctuation                   
can be best expressed via the charge density 
in $^{208}{\rm Pb}$ as
 \begin{equation}
\delta\rho =\rho_{C}(1.8\;\; {\rm fm}) - \rho_{C}(0.0\;\; {\rm fm}).
 \end{equation}
The empirical value is $-0.0023$ ${\rm fm}^{-3}$ \cite{Rei86}, 
which is nicely reproduced
in the RHA (see Table II) while the RMF overestimates $\delta\rho$ by
a factor of 3, sharing the same disease with the non-relativistic mean field
theory \cite{Fri86}.

The effects of the Dirac sea on the physical quantity, specifically, the binding
energy per nucleon are displayed in Table III for eight spherical nuclei. The
calculations are performed with the RHA1 set of parameters. It can be found that
the theoretical predictions including the vacuum contributions are in good agreement
with the empirical values. The magnitudes of the Dirac-sea corrections are quite
similar for different nuclei. The absolute ratios of the Dirac-sea effects and the 
binding energy stay between 16\% and 17\%. All corrections decrease the binding energy
per nucleon. The contributions of the derivative terms are in the same order 
of magnitude as the
effective potential terms. When the nucleus becomes heavier the derivative terms turn 
out to be smaller since the fields become more stable. In all cases, the effects of 
the one-nucleon-loop is about four to five times of that of the one-meson-loop.

Now let us go to the single-particle levels.
In Table IV and V  we present the results of both positive- and negative-energy
proton and neutron
spectra of $^{16}{\rm O}$, $^{40}{\rm Ca}$ and $^{208}{\rm Pb}$. The 
binding energy per nucleon and the {\em rms} charge radius are given too.   
The numerical calculations are performed within two frameworks, i.e., the RHA
including the contributions of the negative-energy sector to the source terms
of the meson fields and the RMF taking into account
 only the valence nucleons as the 
meson-field sources. The experimental data are taken from Ref. \cite{Mat65}.
From the table one can see that all four sets of parameters can reproduce
the empirical values of the binding energies, the {\em rms} charge radii
and the single-particle energies of positive-energy states fairly well. 
For the $E/A$
and the $r_{ch}$, the agreement between the theoretical predictions and the
experimental data are improved from the LIN to the RHA0, RHA1 and NL1 set
of parameters. For the spectra of positive-energy states, due to large
error bars, it seems to be difficult to queue up 
the different sets of parameters. However,
because of the large effective nucleon mass, in the current models
the RHA has smaller spin-orbit splitting (see $1p_{1/2}$ and $1p_{3/2}$ 
state) compared to the RMF. This situation can be improved through introducing 
a tensor coupling for the $\omega$ meson \cite{Rei86}.             
With a suitable chosen coupling strength for the tensor term, a reasonable
spin-orbit splitting may be obtained while a large $m^{*}$ remains.
The effects of the tensor-coupling terms will be investigated in the future studies.
For the negative-energy sector, no experimental
data are available. In all four cases, the potentials of negative-energy 
nucleons are much deeper than the potentials of positive-energy nucleons.
On the other hand, 
one can notice the drastic difference between the RHA and
the RMF calculations -- the single-particle energies calculated from the RHA 
are about half of that from the RMF as can be expected from Fig. 2 and 3,  
exhibiting the importance of taking into account the
Dirac sea effects. It demonstrates that the negative-energy spectra deserve
a sensitive probe to the effective interactions in addition to the positive-energy
spectra.
The spin-orbit splitting of negative-energy sector is so
small that one nearly can not distinguish the $1\bar{p}_{1/2}$ and the 
$1\bar{p}_{3/2}$ state. This is because the spin-orbit potential is related to
$d(S+V)/dr$ in the negative-energy sector and two fields cancel each 
other to a large extent. 
Nevertheless, the space between the $1\bar{s}$ and the $1\bar{p}$ state is
still evident, especially for lighter nuclei. This might be helpful to 
separate the process of knocking out a $1\bar{s}_{1/2}$ negative-energy
nucleon from the background -- a promising way to measure the potential of
the anti-nucleon in laboratory.

Fig.~7 depicts the potentials of the anti-nucleon 
in $^{208}{\rm Pb}$ computed within the NL1 and
the RHA1 model. The corresponding potentials of the nucleon are presented too.
One can easily find that  the NL1 and the RHA1 set of parameters give the
similar potentials for the nucleon (except  in the center of the nucleus
where the results of the RHA1 model exhibit certain fluctuations caused by the
$\rho_{S}^{sea}$ contributed from the vacuum) while the potentials of the anti-nucleon
differ substantially. A much weaker anti-nucleon potential is obtained in the
RHA1 model. This results in a large critical density (around 9.5$\rho_{0}$)
for the spontaneous nucleon--anti-nucleon pair creation as can be seen in Fig.~8
where the density dependence of the nucleon and the anti-nucleon energy 
in symmetric nuclear matter is given. It should be mentioned that the results of 
Fig.~7 and 8 are very sensitive to the effective nucleon mass which          
 can not be determined unambiguously in a model-independent way via experiments.
 In order to know the individual scalar and vector potentials, one has to analyse
 the empirical data both from the positive-energy and the negative-energy sector.
 The later is, unfortunately, currently unavailable. Further investigation is
 apparently needed before coming to a definite conclusion.

\end{sloppypar}

\begin{center}
{\bf VI. SUMMARY AND OUTLOOK}
\end{center}
\begin{sloppypar}
This paper develops a model to describe the bound states of positive energy and 
negative energy in nuclei. Instead of expanding the field operator of the Dirac 
equation in plane waves of nucleons with positive energy and negative
energy, here the expansion is done with plane waves for 
nucleons and anti-nucleons separately, 
both of them with positive energy. In this case, the Dirac equations of both, 
the nucleon and the 
anti-nucleon, can be reduced to Schr\"{o}dinger-equivalent equations.
For each equation, we then search for only one solution, i.e., the 
positive-energy solution.  The numerical procedure 
 is similar to the one
currently used in the relativistic mean field theory for the bound states of 
positive energy, except that one more equation for the anti-nucleon is 
implemented.
Thus, the model is solvable with existing techniques. 
The single-particle energy of the
nucleon with negative energy is just the negative of the single-particle
energy of the anti-nucleon with positive energy, i.e., the solution of the
anti-nucleon's Schr\"{o}dinger equation. 

The contributions of the Dirac sea to the source terms of meson fields can be
separated into two parts, that is, the contributions of the negative-energy
bound states    and the negative-energy continuum. In principle, one should 
use the wave functions of anti-nucleons to evaluate the effects of the bound
states emerging from the Dirac sea. There might exist a large number
of anti-nucleons in  bound states. Hence, it is apparently not 
practical to calculate
all those wave functions in the iteration procedure. 
Here we employ the technique of derivative expansion
for the one-meson and one-nucleon loop to take into account the vacuum 
contributions of both the bound states and the continuum.

In the numerical procedure, we first fit the parameters of the model to the 
properties of spherical nuclei. Two sets of parameters with and without
nonlinear self-interaction of the scalar field are obtained for the 
present RHA model including the vacuum contributions. They are then used 
to investigate the vacuum polarization effects on both positive-energy and
negative-energy sector. The corresponding calculations of the RMF model are also
presented for comparison. Our results show that both the RMF and the RHA model
describe the properties of spherical nuclei very well. Due to the feedback of the
vacuum to the meson fields, the scalar and vector potentials decrease in the 
RHA. This causes the drastic difference on the single-particle energies
of negative-energy nucleons calculated within the RHA model and within the
RMF model, while the single-particle
energies of positive-energy nucleons coincide each other within two models. 
Since the negative-energy sector is sensitive to the sum of the scalar and 
vector field $-V+S$ while the positive-energy sector is sensitive to the
cancellation of the fields $V+S$, the study of both of them in an unified 
framework will lead to the determination of  the individual $S$
and $V$! Thus, 
 it is currently very
important to have experimental data to check the theoretical predicted bound
levels of negative energy. 
It will provide us with a chance to judge the physical necessity of introducing
strong scalar and vector potentials in the Dirac picture.
If this picture  is valid
for the nucleon-nucleus and anti-nucleon--nucleus interactions, 
a fascinating direction of future studies
is to investigate the vacuum correlation and the collective production of 
the anti-nuclei in relativistic heavy-ion collisions. Experimental efforts
in this direction are presently underway \cite{Ars99}.

\end{sloppypar}
 \begin{center}
{\bf ACKNOWLEDGMENTS}
 \end{center}
 \begin{sloppypar}
The authors thank P.-G.~Reinhard, Zhongzhou Ren, J. Schaffner-Bielich
and C.~Beckmann for stimulating discussions.  
 G.~Mao gratefully acknowledges  the Alexander von Humboldt-Stiftung
 for financial support      
 and  the people at the
Institut f\"{u}r
Theoretische Physik der J.~W.~Goethe Universit\"{a}t for their hospitality.
This work was supported by DFG-Graduiertenkolleg Theoretische und
Experimentelle
Schwerionenphysik, GSI, BMBF, DFG and A.v. Humboldt-Stiftung.                             

 \end{sloppypar}

\newpage
\def\baselinestretch{1.0}

\begin{table}
\caption{The experimental values for the observables included in the fit,
the binding energy $E_{B}$, diffraction radius $R$ and surface thickness
$\sigma$. In the last line we also give the adopted errors $\Delta {\rm O}_{n}$
for the fit.} \vspace{0.5cm}
\begin{tabular}{lcccc}
                    &   $E_{B}$ (MeV)   &  $R$ (fm)   &  $\sigma$ (fm)  \\
 \hline
$^{16}{\rm O}$    &  $-$127.6  & 2.777       &   0.839   \\           
$^{40}{\rm Ca}$   &  $-$342.1  & 3.845       &   0.978   \\   
$^{48}{\rm Ca}$   &  $-$416.0  & 3.964       &   0.881   \\   
$^{58}{\rm Ni}$   &  $-$506.5  & 4.356       &   0.911   \\         
$^{90}{\rm Zr}$   &  $-$783.9  & 5.040       &   0.957   \\   
$^{116}{\rm Sn}$  &  $-$988.7  & 5.537       &   0.947   \\   
$^{124}{\rm Sn}$  &  $-$1050.0 & 5.640       &   0.908   \\   
$^{208}{\rm Pb}$  &  $-$1636.4 & 6.806       &   0.900   \\    \\
$\Delta {\rm O}_{n} / {\rm O}_{n}$  &  0.2\%  &  0.5\%   &   1.5\%

\end{tabular}
\end{table}

\begin{table}
\caption{Parameters of the RMF and the RHA models as well as the corresponding 
saturation properties. The results of shell fluctuation and 
the $\chi^{2}$ values of different sets of parameters are
also presented.} \vspace{0.5cm}
\begin{tabular}{lcccc}
 & \multicolumn{2}{c}{RMF} & \multicolumn{2}{c}{RHA}    \\
 & LIN & NL1 & RHA0 & RHA1  \\
 \hline
$M_{N}$ (MeV)        &  938.000    &  938.000 &  938.000 &   938.000   \\
$m_{\sigma}$ (MeV)   &  615.000    &  492.250 &  615.000 &   458.000   \\
$m_{\omega}$ (MeV)   &  1008.00    &  795.359 &  916.502 &   816.508   \\
$m_{\rho}$ (MeV)     &  763.000    &  763.000 &  763.000 &   763.000   \\
${\rm g}_{\sigma}$   &  12.3342    &  10.1377 &  9.9362  &   7.1031    \\
${\rm g}_{\omega}$   &  17.6188    &  13.2846 &  11.8188 &   8.8496    \\
${\rm g}_{\rho}$     &  10.3782    &  9.9514  &  10.0254 &   10.2070   \\
$b$ (fm$^{-1}$)      &  0.0        &  24.3448 &  0.0     &  24.0870    \\
$c$                  &  0.0        &$-$217.5876&  0.0     & $-$15.9936  \\
\\
$\rho_{0}$ (fm$^{-3}$)& 0.1525     &  0.1518  &  0.1513  &   0.1524    \\
$E/A$ (MeV)           & $-$17.03   &  $-$16.43& $-$17.39 &  $-$16.98   \\
$m^{*}/M_{N}$         & 0.533      &  0.572   & 0.725    &   0.788     \\
$K$ (MeV)             & 580        &  212     &  480     &   294       \\
$a_{4}$ (MeV)         & 46.8       &  43.6    &  40.4    &   40.4      \\
\\
$\delta\rho$ in $^{208}{\rm Pb}$ (fm$^{-3}$)&
                        $-$0.0075 & $-$0.0070 & $-$0.0016&  $-$0.0030  \\
$\chi ^{2}$              &   1773   &   66      &  1040   & 812

\end{tabular}
\end{table}

\newpage
\begin{table}
\caption{Experimental and theoretical binding energy per nucleon as well as
the vacuum corrections.             
The calculations are performed with the RHA1 set of parameters. The Dirac sea
effects are separated by the effective potential terms and the
derivative terms as well as the contributions of the one-nucleon-loop and
the one-meson-loop.} \vspace{0.5cm}
\begin{tabular}{lccccccc}
& Exp. & Theory & Dirac Sea & Eff. Pot. & Deri. Terms 
                            & Nucl. Loop & Meson Loop \\ 
 \hline
$^{16}{\rm O}$    &  $-$7.98   & $-$8.00    &   1.37   &  0.58   
                  &  0.80      &  1.10       &   0.27   \\
$^{40}{\rm Ca}$   &  $-$8.55   & $-$8.73     &   1.43   & 0.82
                  &  0.60      &  1.17       &   0.26   \\
$^{48}{\rm Ca}$   &  $-$8.67   & $-$8.51     &   1.39   & 0.82
                  &  0.57      &  1.13       &   0.25   \\
$^{58}{\rm Ni}$   &  $-$8.73   & $-$8.44     &   1.44   & 0.87
                  &  0.57      &  1.18       &   0.26   \\
$^{90}{\rm Zr}$   &  $-$8.71   & $-$8.74     &   1.42   & 0.96
                  &  0.46      &  1.16       &   0.25   \\
$^{116}{\rm Sn}$  &  $-$8.52   & $-$8.61     &   1.39   & 0.98
                  &  0.42      &  1.14       &   0.25   \\
$^{124}{\rm Sn}$  &  $-$8.47   &  $-$8.50    &   1.34   & 0.96
                  &  0.39      &  1.10       &   0.24   \\
$^{208}{\rm Pb}$  &  $-$7.87   & $-$7.93     &   1.30   & 0.98
                  &  0.32      &  1.07       &   0.23

\end{tabular}
\end{table}

\newpage
\begin{table}
\caption{The single-particle energies of both positive- and negative-energy 
protons as well as the binding energy per nucleon and the {\em rms} charge
radius in $^{16}{\rm O}$, $^{40}{\rm Ca}$ and
 $^{208}{\rm Pb}$. } \vspace{0.5cm}
\begin{tabular}{cccccc}
 & \multicolumn{2}{c}{RMF} & \multicolumn{2}{c}{RHA} &   \\
 & LIN & NL1 & RHA0 & RHA1 & EXP. \\
 \hline
$\;\;\;\;\;\;\;\;^{16}{\rm O}$ &       &          &           &      &       \\
$E/A$ (MeV)   &   7.80   &   8.00   &    8.01   & 8.00 &   7.98 \\
$r_{ch}$ (fm) &   2.59   &   2.73   &    2.62   & 2.66 &   2.74 \\
POS. ENE.     &          &          &           &      &        \\
$1s_{1/2}$ (MeV)& 42.99  &  36.18   &   32.21   & 30.68&   40$\pm$8 \\
$1p_{3/2}$ (MeV)& 20.71  &  17.31   &   16.09   & 15.23&   18.4   \\
$1p_{1/2}$ (MeV)& 10.85  &  11.32   &   12.98   & 13.24&   12.1  \\
NEG. ENE.     &          &          &           &      &        \\
$1\bar{s}_{1/2}$ (MeV)& 821.30 &  674.11  &   413.62  &299.42&            \\
$1\bar{p}_{3/2}$ (MeV)& 754.62 &  604.70  &   369.78  &258.40&          \\
$1\bar{p}_{1/2}$ (MeV)& 755.43 &  605.77  &   370.36  &258.93&         \\
\hline
$\;\;\;\;\;\;\;^{40}{\rm Ca}$ &       &          &           &      &       \\
$E/A$ (MeV)   &   8.38   &   8.58   &    8.65   & 8.73 &   8.55 \\
$r_{ch}$ (fm) &   3.36   &   3.48   &    3.39   & 3.42 &   3.45 \\
POS. ENE.     &          &          &           &      &        \\
$1s_{1/2}$ (MeV)& 51.21  &  46.86   &   38.64   & 36.58&  50$\pm$11\\
$1p_{3/2}$ (MeV)& 35.05  &  30.15   &   27.11   & 25.32&          \\
$1p_{1/2}$ (MeV)& 29.25  &  25.11   &   25.17   & 24.03&  34$\pm$6  \\
NEG. ENE.     &          &          &           &      &        \\
$1\bar{s}_{1/2}$ (MeV)& 840.76 &  796.09  &   456.58  &339.83&            \\
$1\bar{p}_{3/2}$ (MeV)& 792.36 &  706.36  &   424.85  &309.24&          \\
$1\bar{p}_{1/2}$ (MeV)& 792.75 &  707.86  &   425.14  &309.52&         \\
\hline
$\;\;\;\;\;\;\;^{208}{\rm Pb}$ &       &          &           &      &       \\
$E/A$ (MeV)   &   7.83   &   7.89   &    7.96   & 7.93 &   7.87 \\
$r_{ch}$ (fm) &   5.34   &   5.52   &    5.43   & 5.49 &   5.50 \\
POS. ENE.     &          &          &           &      &        \\
$1s_{1/2}$ (MeV)& 58.71  &  50.41   &   44.43   &40.80 &           \\
$1p_{3/2}$ (MeV)& 52.74  &  44.45   &   39.87   &36.45 &          \\
$1p_{1/2}$ (MeV)& 51.83  &  43.75   &   39.49   &36.21 &            \\
NEG. ENE.     &          &          &           &      &        \\
$1\bar{s}_{1/2}$ (MeV)& 830.16 &  717.01  &   476.61  &354.18&            \\
$1\bar{p}_{3/2}$ (MeV)& 819.15 &  705.20  &   466.08  &344.48&          \\
$1\bar{p}_{1/2}$ (MeV)& 819.22 &  705.28  &   466.13  &344.52&           

\end{tabular}
\end{table}

\newpage
\begin{table}
\caption{The single-particle energies of both positive- and negative-energy
neutrons.} \vspace{0.5cm}
\begin{tabular}{cccccc}
 & \multicolumn{2}{c}{RMF} & \multicolumn{2}{c}{RHA} &   \\
 & LIN & NL1 & RHA0 & RHA1 & EXP. \\
 \hline
$\;\;\;\;\;\;\;\;^{16}{\rm O}$ &       &          &           &      &       \\
POS. ENE.     &          &          &           &      &        \\
$1s_{1/2}$ (MeV)& 47.23  &  40.21   &   36.33   & 34.71&   45.7  \\
$1p_{3/2}$ (MeV)& 24.70  &  21.07   &   19.99   & 19.04&   21.8   \\
$1p_{1/2}$ (MeV)& 14.74  &  15.01   &   16.86   & 17.05&   15.7  \\
NEG. ENE.     &          &          &           &      &        \\
$1\bar{s}_{1/2}$ (MeV)& 814.98 &  667.93  &   407.44  &293.23&            \\
$1\bar{p}_{3/2}$ (MeV)& 748.48 &  598.74  &   363.83  &252.48&          \\
$1\bar{p}_{1/2}$ (MeV)& 749.22 &  599.74  &   364.37  &252.97&         \\
\hline
$\;\;\;\;\;\;\;^{40}{\rm Ca}$ &       &          &           &      &       \\
POS. ENE.     &          &          &           &      &        \\
$1s_{1/2}$ (MeV)& 59.36  &  54.85   &   46.67   & 44.48&         \\
$1p_{3/2}$ (MeV)& 42.94  &  37.79   &   34.91   & 32.98&          \\
$1p_{1/2}$ (MeV)& 37.17  &  32.73   &   32.99   & 31.71&           \\
NEG. ENE.     &          &          &           &      &        \\
$1\bar{s}_{1/2}$ (MeV)& 828.82 &  783.87  &   444.85  &327.96&            \\
$1\bar{p}_{3/2}$ (MeV)& 781.18 &  694.80  &   413.71  &298.04&          \\
$1\bar{p}_{1/2}$ (MeV)& 781.46 &  696.18  &   413.93  &298.26&         \\
\hline
$\;\;\;\;\;\;\;^{208}{\rm Pb}$ &       &          &           &      &       \\
POS. ENE.     &          &          &           &      &        \\
$1s_{1/2}$ (MeV)& 65.19  &  58.97   &   50.99   & 47.40&           \\
$1p_{3/2}$ (MeV)& 58.50  &  52.44   &   46.05   & 42.66&          \\
$1p_{1/2}$ (MeV)& 57.73  &  51.82   &   45.71   & 42.45&            \\
NEG. ENE.     &          &          &           &      &        \\
$1\bar{s}_{1/2}$ (MeV)& 789.37 &  678.23  &   435.30  &313.18&            \\
$1\bar{p}_{3/2}$ (MeV)& 779.44 &  667.70  &   425.81  &304.61&          \\
$1\bar{p}_{1/2}$ (MeV)& 779.45 &  667.73  &   425.82  &304.61&           

\end{tabular}
\end{table}

 \newpage
 \begin{center}
 {\bf FIGURE CAPTIONS}
 \end{center}
\def\baselinestretch{1.5}
 \begin{description}
 \item[\tt Fig.1 ] Equations of state of nuclear matter as well as the 
 scalar and vector potentials calculated within the RMF and the RHA models
 under different sets of parameters given in Table II.
 \item[\tt Fig.2 ] The scalar potentials in $^{16}{\rm O}$, $^{40}{\rm Ca}$
 and $^{208}{\rm Pb}$.
 \item[\tt Fig.3 ] The vector potentials in $^{16}{\rm O}$, $^{40}{\rm Ca}$
 and $^{208}{\rm Pb}$.
 \item[\tt Fig.4 ] The fractions of the baryon density contributed from the
 Dirac sea. The calculations are performed for $^{16}{\rm O}$,
 $^{40}{\rm Ca}$ and $^{208}{\rm Pb}$ with the RHA1 set of parameters.
 \item[\tt Fig.5 ] The upper panel displays the baryon density in $^{16}{\rm O}$
 computed with the NL1 and the RHA1 set of parameters, respectively. 
 The lower panel shows the contributions of the valence nucleons and the Dirac sea
 to the baryon density. The calculations are performed for $^{16}{\rm O}$
 with the RHA1 set of parameters.
 \item[\tt Fig.6 ] The charge densities in $^{16}{\rm O}$, $^{40}{\rm Ca}$
 and $^{208}{\rm Pb}$.
 \item[\tt Fig.7 ] The potentials of the nucleon and the anti-nucleon 
 in $^{208}{\rm Pb}$.
 \item[\tt Fig.8 ] The single-particle energies of the nucleon and the anti-nucleon
 as a function of density. The critical point for the NL1 is around 
 3.3 $\rho_{0}$ while for the RHA1 is around 9.5 $\rho_{0}$.
  \end{description}
 \newpage
 {\Large Fig. 1}
 \begin{figure}[htbp]
  \vspace{0cm}
 \hskip  1cm \psfig{file=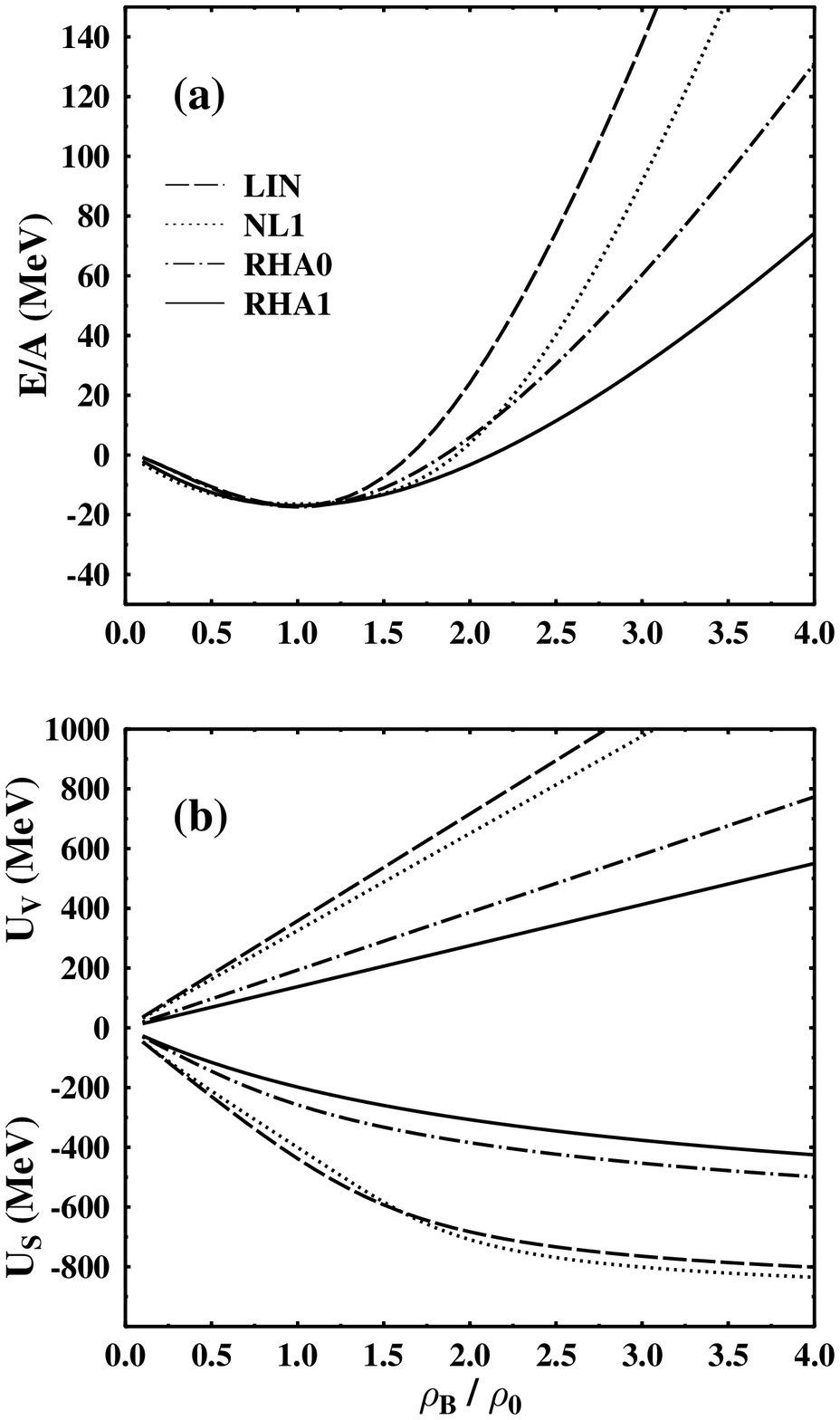,width=12.cm,height=19cm,angle=0}
\end{figure}
 \newpage
 {\Large Fig. 2}
 \begin{figure}[htbp]
  \vspace{0cm}
 \hskip  -1cm \psfig{file=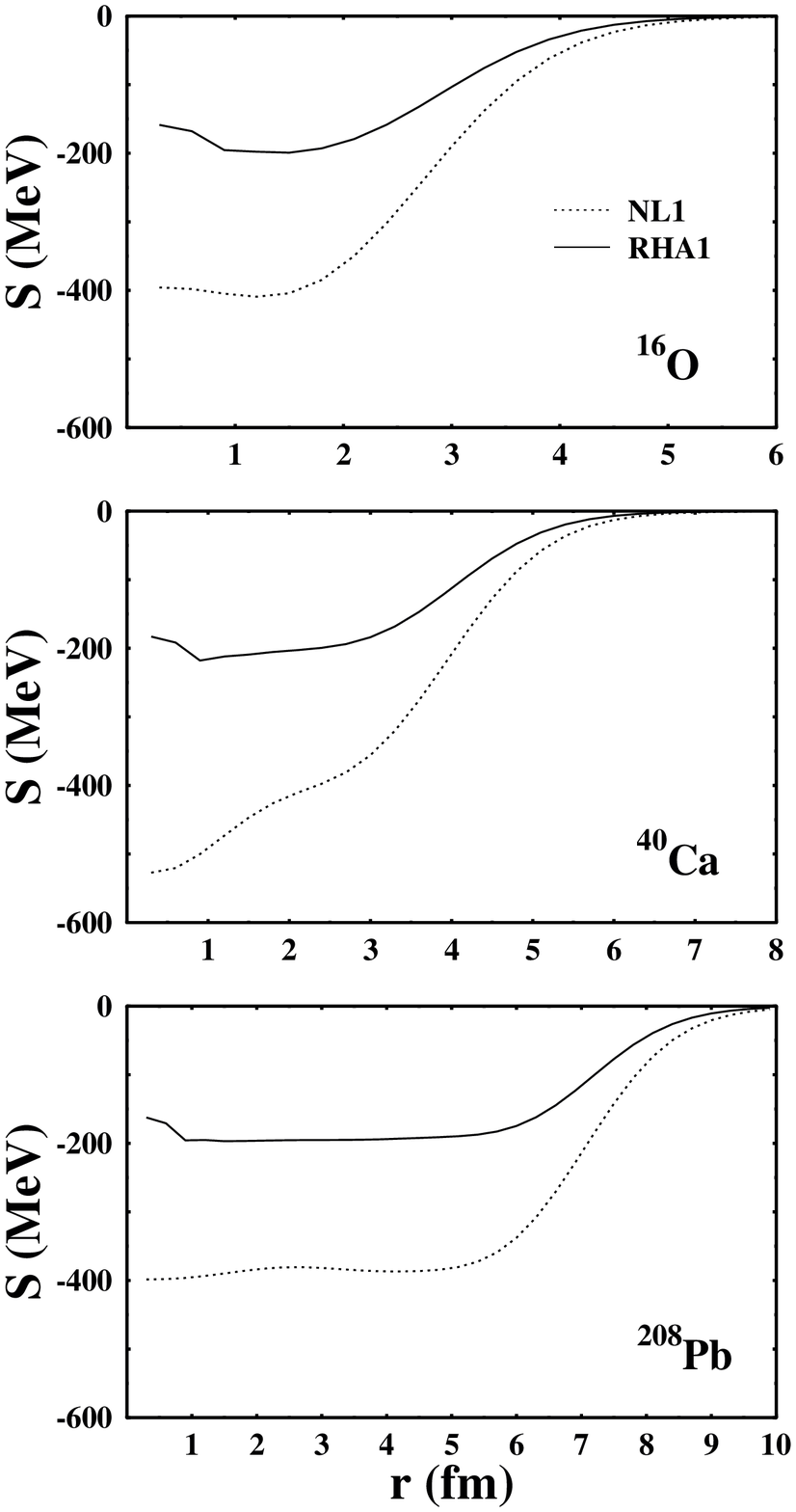,width=15.cm,height=21cm,angle=0}
\end{figure}
 \newpage
 {\Large Fig. 3}
 \begin{figure}[htbp]
  \vspace{0cm}
 \hskip  -1cm \psfig{file=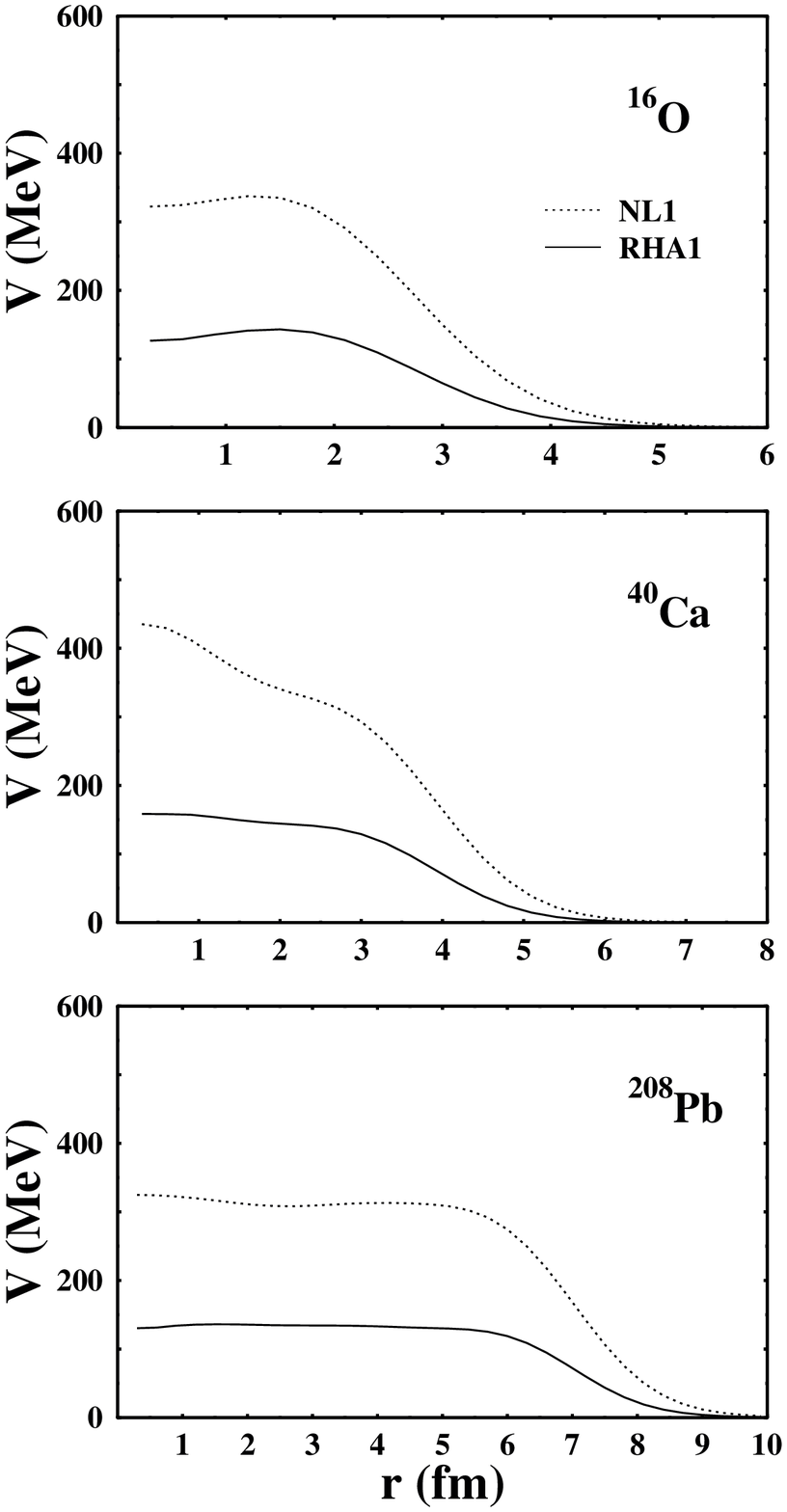,width=15.cm,height=21cm,angle=0}
\end{figure}
 \newpage
 {\Large Fig. 4}
 \begin{figure}[htbp]
  \vspace{0cm}
 \hskip  -1cm \psfig{file=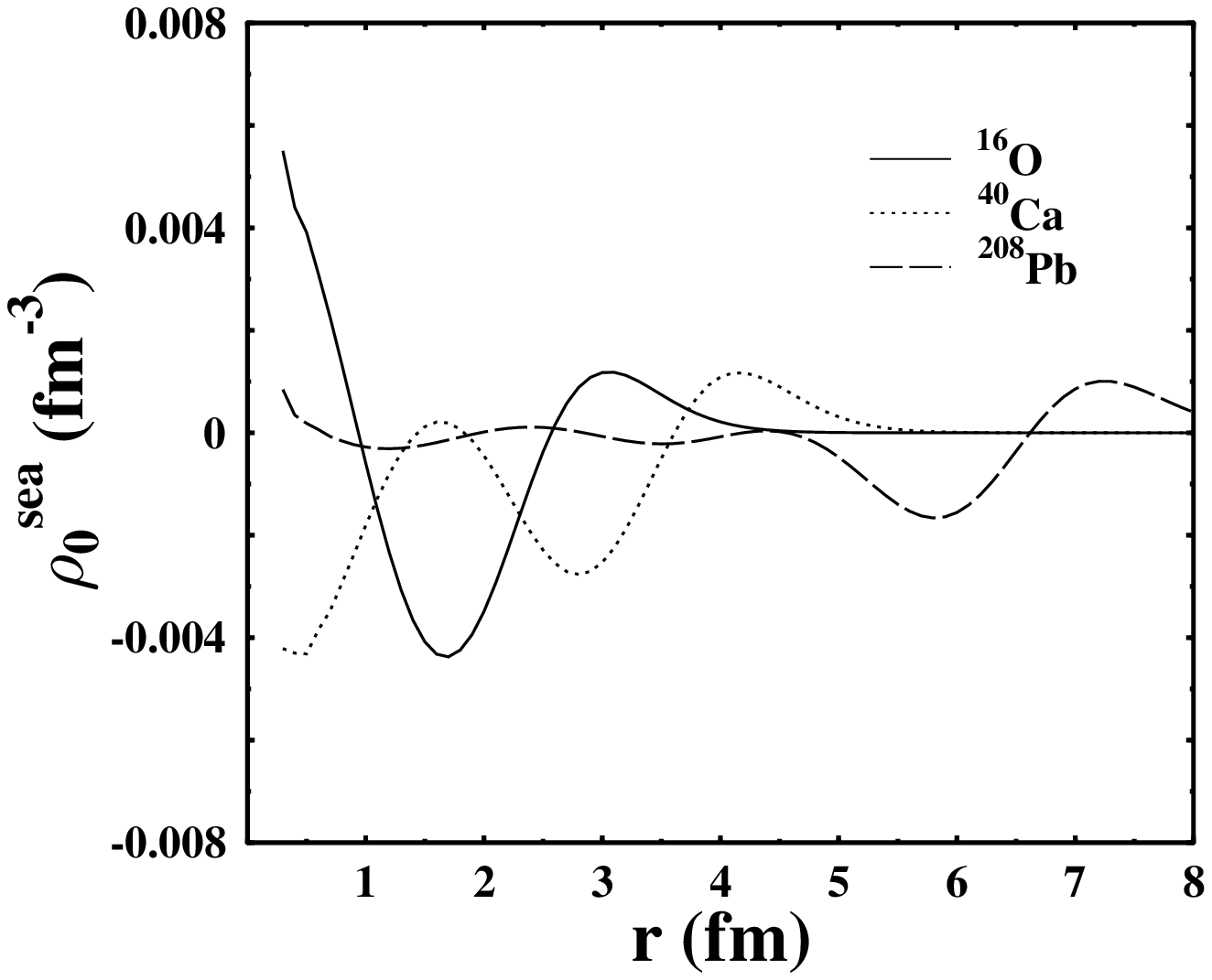,width=12.cm,height=15cm,angle=-90}
\end{figure}
 \newpage
 {\Large Fig. 5}
 \begin{figure}[htbp]
  \vspace{0cm}
 \hskip  -1cm \psfig{file=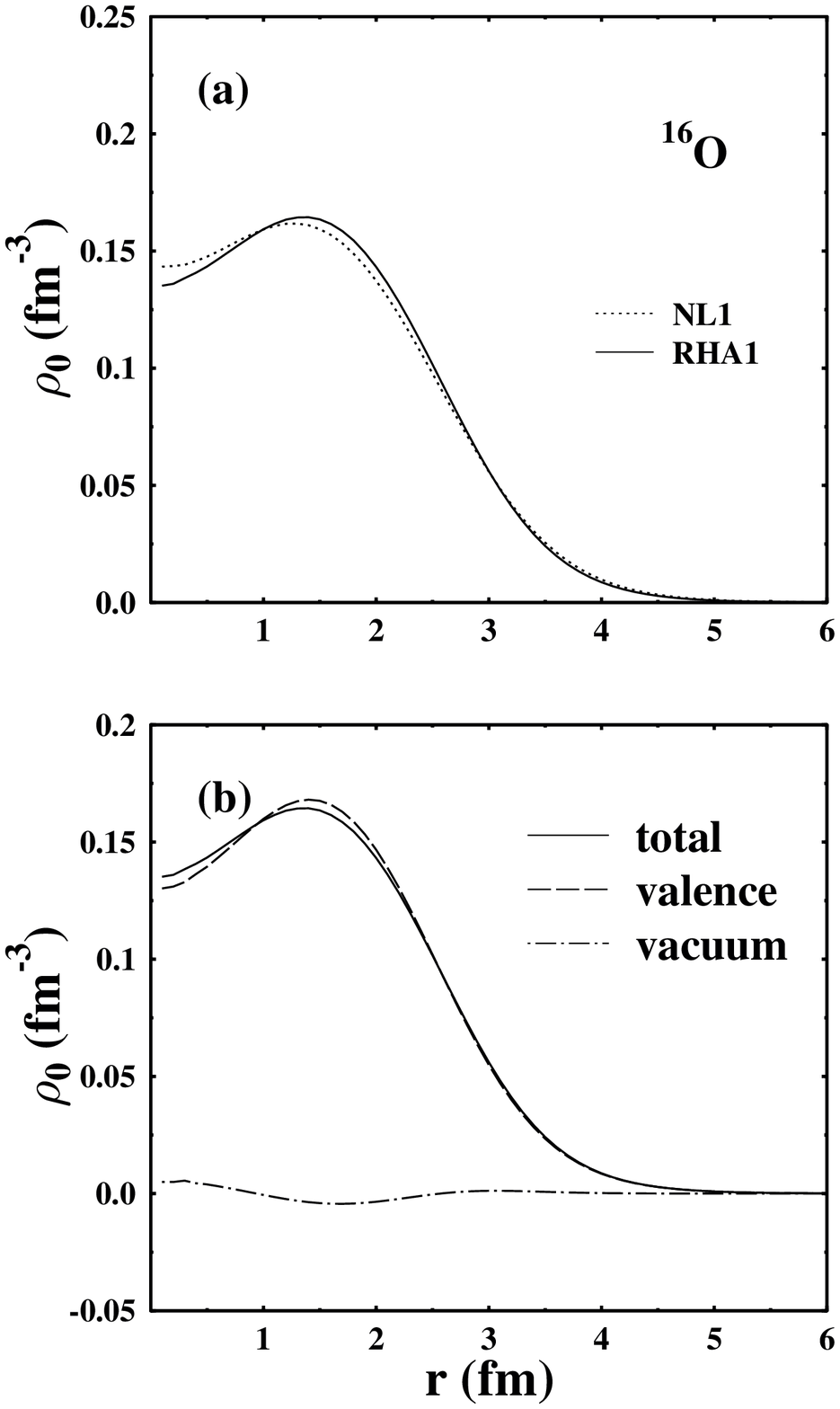,width=12.cm,height=19cm,angle=0}
\end{figure}
 \newpage
 {\Large Fig. 6}
 \begin{figure}[htbp]
  \vspace{0cm}
 \hskip  -2cm \psfig{file=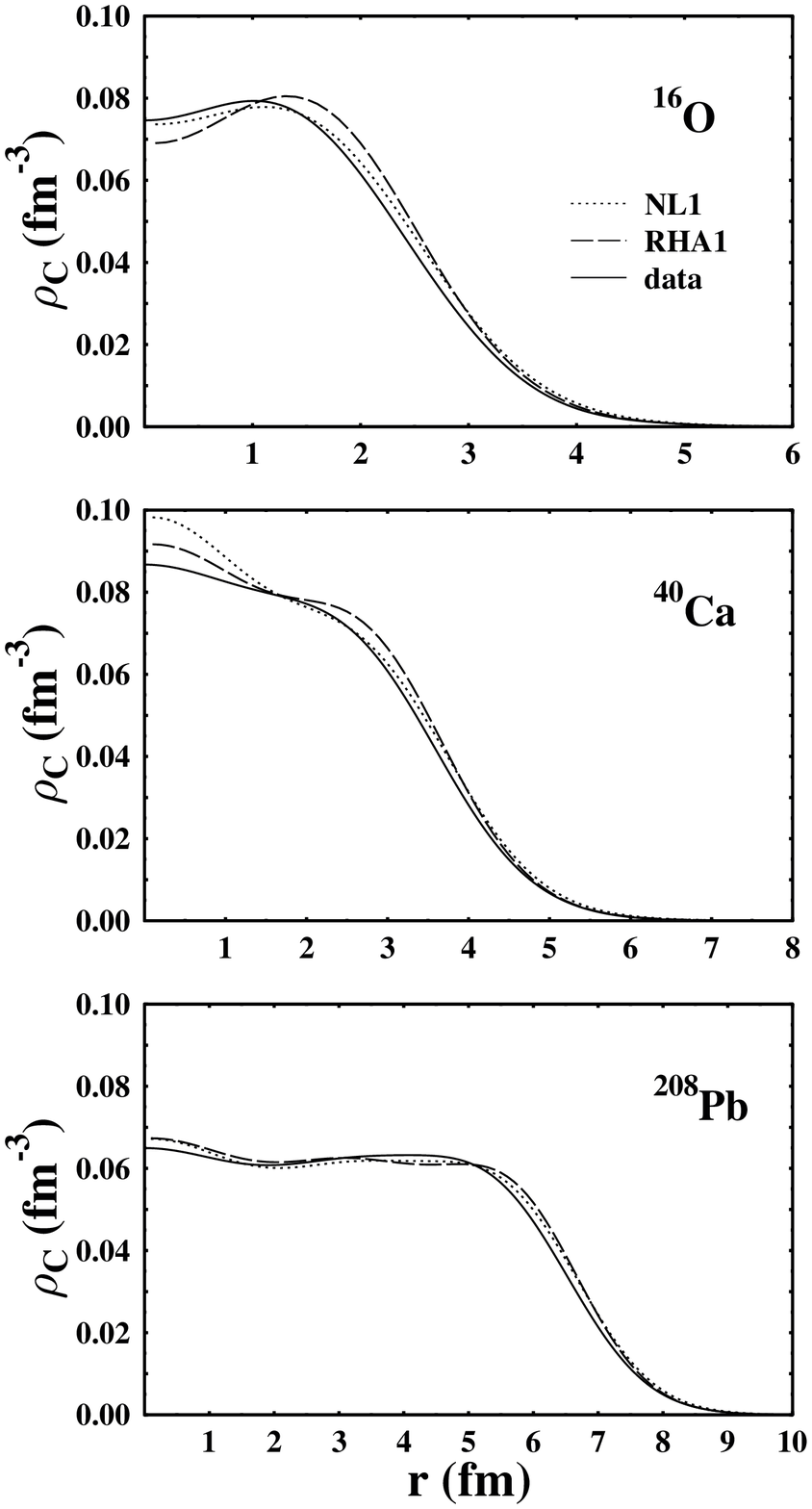,width=15.cm,height=21cm,angle=0}
\end{figure}
 \newpage
 {\Large Fig. 7}
 \begin{figure}[htbp]
  \vspace{0cm}
 \hskip  -1cm \psfig{file=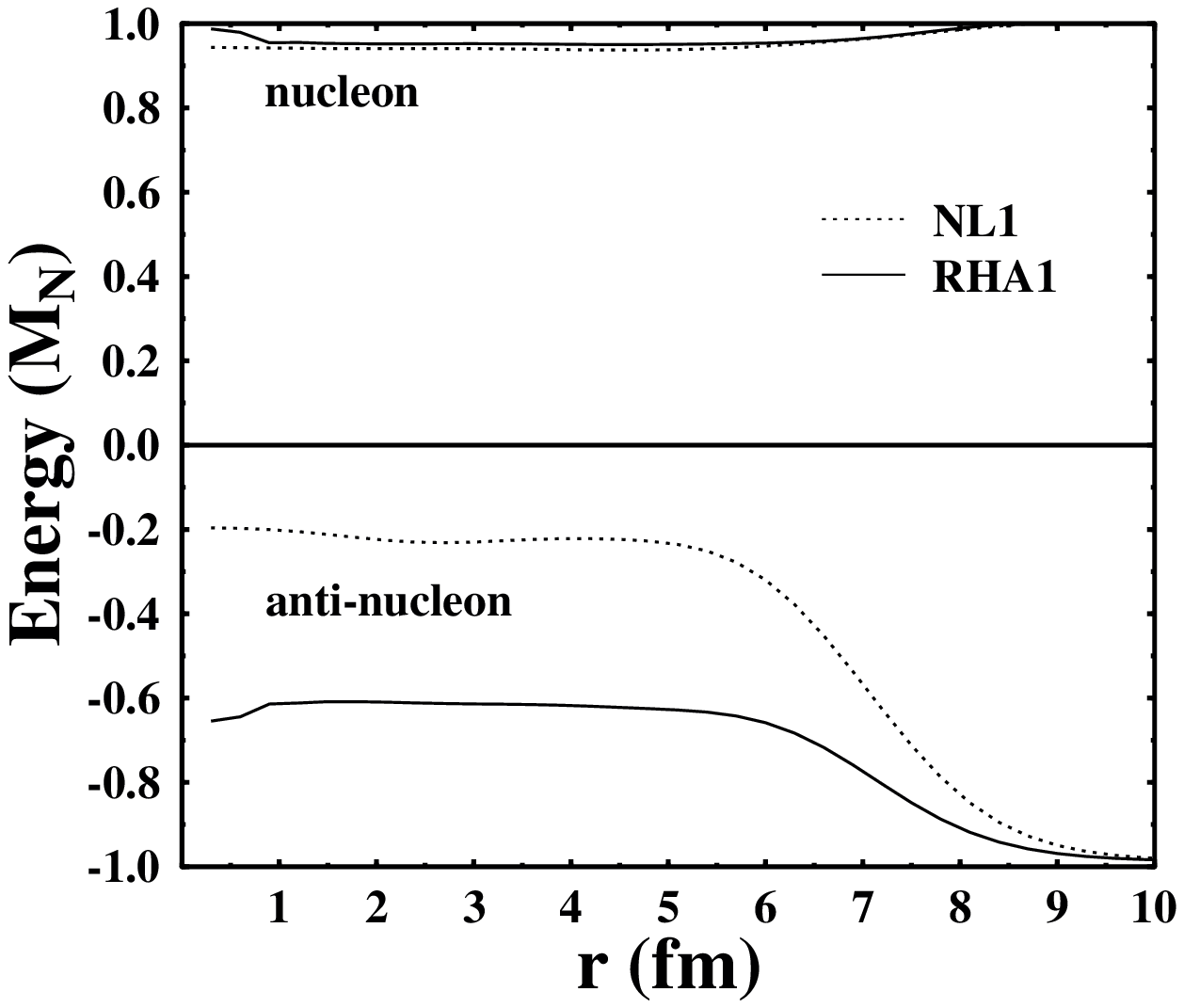,width=12.cm,height=15cm,angle=-90}
\end{figure}
 \newpage
 {\Large Fig. 8}
 \begin{figure}[htbp]
  \vspace{0cm}
 \hskip  -1cm \psfig{file=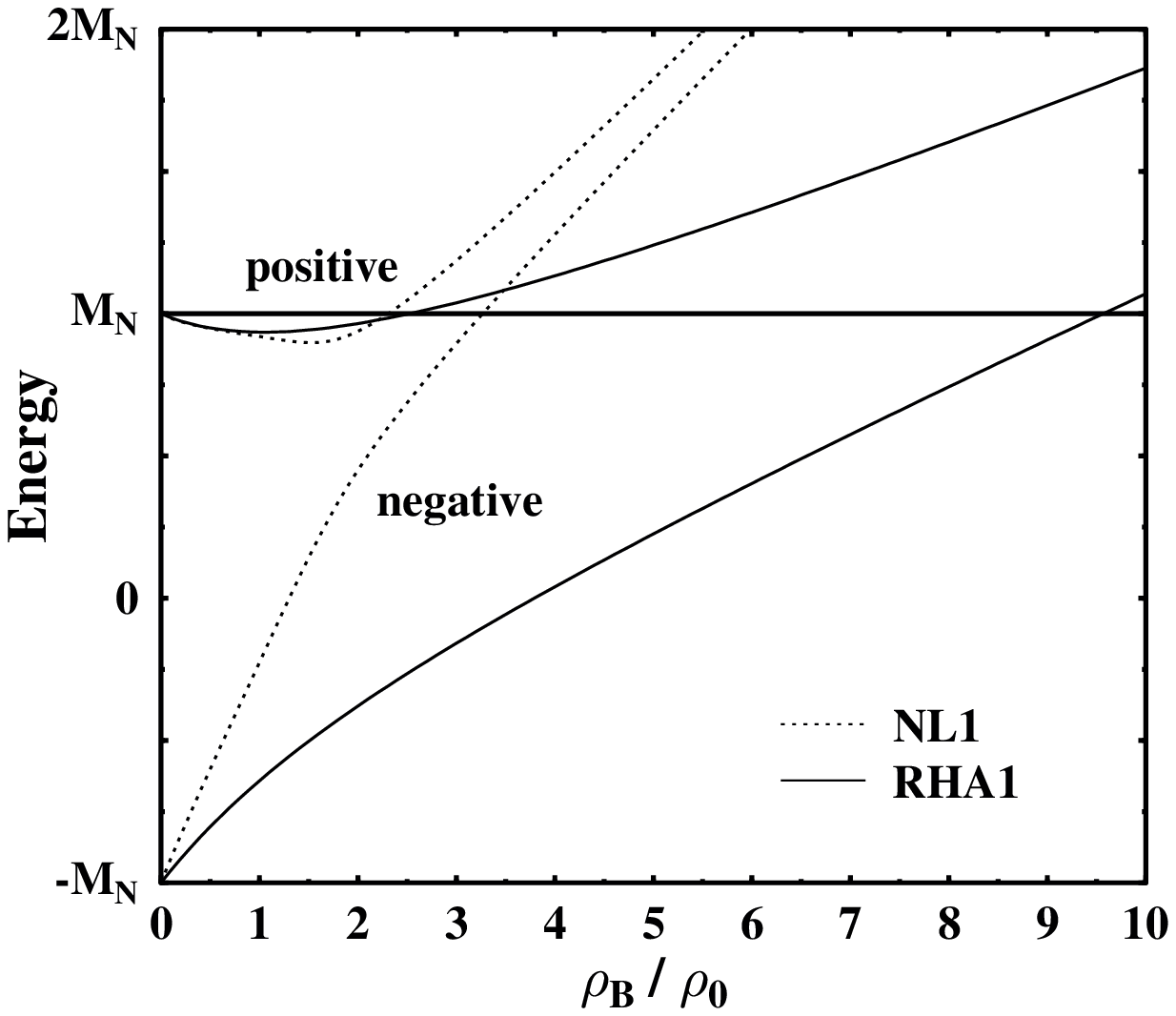,width=12.cm,height=15cm,angle=-90}
\end{figure}

\end{document}